\documentclass[twocolumn,showpacs,preprintnumbers,%
amsmath,amssymb,superscriptaddress,aps,floatfix]{revtex4}
\usepackage{graphicx}% Include figure files
\usepackage{dcolumn}% Align table columns on decimal point
\usepackage{bm}% bold math
\usepackage{times}
\usepackage{mathbbol}
\def\onehalf{\textstyle\frac{1}{2}}
\def\quarter{\textstyle\frac{1}{4}}
\def\onethird{\textstyle\frac{1}{3}}
\def\vecq{\bm{q}}
\def\vecp{\bm{p}}
\def\vecs{\bm{s}}
\def\bxi{\bm{\xi}}
\def\bfeta{\bm{\eta}}
\makeatletter
\def\Tr{\mathop{\operator@font Tr}\nolimits}
\makeatother
\def\d{d}
\def\HC{h}
\def\text#1{\mathrm{#1}}
\newtheorem{adri}{Theorem}
\begin{document}
\title{Energy-second-moment map analysis as an approach
to quantify the irregularity\\ of Hamiltonian systems}
\author{J\"urgen Struckmeier}
\email{j.struckmeier@gsi.de}
\affiliation{Gesellschaft f\"ur Schwerionenforschung (GSI),
Planckstr.~1, D-64291~Darmstadt, Germany}
\affiliation{Institut f\"ur Angewandte Physik der
Johann Wolfgang Goethe-Universit\"at,
Max-von-Laue-Str.~1, D-60438~Frankfurt am Main, Germany}
\author{Andreas Redelbach}
\affiliation{Gesellschaft f\"ur Schwerionenforschung (GSI),
Planckstr.~1, 64291~Darmstadt, Germany}
\received{23 March 2004; revised manuscript received 27 February 2006;
published 21 August 2006}
\begin{abstract}
A different approach will be presented that aims to scrutinize the
phase-space trajectories of a general class of Hamiltonian
systems with regard to their regular or irregular behavior.
The approach is based on the `energy-second-moment map' that can
be constructed for all Hamiltonian systems of the generic form
$H=\vecp^{2}/2+V(\vecq,t)$.
With a three-component vector $\vecs$ consisting of the system's
energy $\HC$ and second moments $\vecq\vecp$, $\vecq^{2}$,
this map linearly relates the vector $\vecs(t)$ at time $t$
with the vector's initial state $\vecs(0)$ at $t=0$.
It will turn out that this map is directly obtained from
the solution of a linear third-order equation that
establishes an extension of the set of canonical equations.
The Lyapunov functions of the energy-second-moment map
will be shown to have simple analytical representations in
terms of the solutions of this linear third-order equation.
Applying Lyapunov's regularity analysis for linear systems,
we will show that the Lyapunov functions of the
energy-second-moment map yields information on the
irregularity of the particular phase-space trajectory.
Our results will be illustrated by means of numerical examples.
\end{abstract}
\maketitle
\section{Introduction}
The phase-space trajectories of dynamical systems can be
classified as being either regular or irregular.
The distinction between regular and irregular behavior of a
trajectory of a given $n$-degree-of-freedom
dynamical system is commonly made on the basis of the
linear system of perturbation equations that can be solved
in conjunction with the full set of equations of motion.
The $n\times n$ solution matrix of the system of perturbation
equations is referred to as the stability matrix.
This matrix thus describes the stability of all trajectories
with respect to small changes of their initial values.
From the stability matrix, the system's degree of irregularity
can then be quantified in terms of Lyapunov's characteristic
exponents (cf., for example, Refs.~\cite{lyapu,mayer,ryabov}).
The ``degree of irregularity'' of a given dynamical system is then
defined as the proportion of the phase-space volume at $t=0$ that
gives rise to irregular trajectories to the corresponding
phase-space volume that leads to regular trajectories.

In this paper, a different approach to quantify the irregularity
of the time evolution of dynamical systems will be presented.
It applies to all---possibly explicitly time-dependent---%
Hamiltonian systems that are canonically equivalent to the
generic Hamiltonian $H(\vecq,\vecp,t)=\vecp^{2}/2+V(\vecq,t)$.

In contrast to conventional approaches that analyze the
stability matrix, our method is based on the analysis
of the ``energy-second-moment map''\cite{struck}.
A derivation of this map will be sketched in Sec.~\ref{sec:esmm}.
Briefly, the energy-second-moment map {\em linearly\/}
relates the three-component vector $\vecs(t)$ of system quantities
energy $\HC$ and second moments $\vecq\vecp$, $\vecq^{2}$ at finite
times $t$ with the vector's initial state $\vecs(0)$ at $t=0$.
Similar to the stability matrix, the energy-second-moment map
is obtained by solving an {\em additional\/} set of linear
differential equations on the top of the set of equations of motion.
It will turn out that this map can be represented by a
$3\times 3$ order matrix $\Xi(t)$ with unit determinant.
As Lyapunov's theory of characteristic coefficients can be
applied to any linear system with time-dependent and
generally nonperiodic coefficients, we may therefore
make use of this theory to analyze the regularity of
the energy-second-moment map.

We will give in Sec.~\ref{sec:rgt} a brief review of the Lyapunov
regularity analysis of nonautonomous, homogeneous linear systems.
Applied to the particular energy-second-moment matrix
$\Xi(t)$, we will show in Sec.~\ref{sec:gtdh} by means of a
$QR$ decomposition of $\Xi(t)$ that its three Lyapunov
functions $\lambda_{1,2,3}(t)$ always have a simple analytical
representation in terms of the column vectors of $\Xi(t)$.
Furthermore, we will see in Sec.~\ref{sec:atih} that our
analysis becomes particularly simple for autonomous
(time-independent) Hamiltonian systems.
On the basis of this astonishing result, we can easily determine
the Lyapunov coefficient of irregularity of the energy-second-moment
map and thereby quantify the system's degree of irregularity.
The regularity analysis emerging from the energy-second-moment
map therefore simplifies the conventional approach that
is based on the analysis of the stability matrix.

By means of several numerical examples, given in Sec.~\ref{sec:ne},
we will illustrate that the degree of irregularity of a
Hamiltonian system can indeed be quantified in terms of the
time evolution of the Lyapunov coefficients $\lambda_{1,2,3}(t)$
of the energy-second-moment map.
We will furthermore show that the time evolution of the Lyapunov
functions of a chaotic trajectory allows to identify particular
time intervals where the particle's motion is quasi-regular.
\section{\label{sec:esmm}Energy-second-moment map}
We consider an $n$-degree-of-freedom system of classical
particles in a $2n$-dimensional Cartesian phase space whose
Hamiltonian can be converted into the generic form
\begin{equation}\label{ham0}
H(\vecq,\vecp,t)=\onehalf\vecp^{2}+V(\vecq,t).
\end{equation}
Herein, $\vecq$ denotes the $n$-dimensional vector of configuration
space variables, and $\vecp$ is the vector of conjugate momenta.
The system's time evolution is then given as the solutions
of the canonical equations
\begin{equation}\label{caneq}
\dot{\vecq}=\vecp\,,\qquad\dot{\vecp}=
-\frac{\partial V(\vecq,t)}{\partial \vecq}.
\end{equation}
Based on the canonical equations~(\ref{caneq}), we can
subsequently set up the equations of motion for the instantaneous
system energy $\HC(t)$, i.e., the {\em value\/} of the Hamiltonian,
\begin{displaymath}
\HC(t)=H(\vecq(t),\vecp(t),t),
\end{displaymath}
and for the second moments $\vecq^{2}(t)$ and $\vecq(t)\vecp(t)$ as
the scalar products of the canonical coordinate vectors $\vecq(t)$
and $\vecp(t)$, respectively,
\begin{equation}\label{exteq}
\frac{\d}{\d t}\HC=\frac{\partial V(\vecq,t)}{\partial t},\;
\frac{\d}{\d t}\vecq^{2}=2\vecq\vecp,\;
\frac{\d}{\d t}\vecq\vecp=\vecp^{2}-
\vecq\,\frac{\partial V(\vecq,t)}{\partial \vecq}.
\end{equation}
Let us now assume that the canonical equations~(\ref{caneq})
were integrated for the given initial conditions, hence that the
actual spatial trajectory $\vecq=\vecq(t)$ is a known function of time.
Then, the potential-related terms $g_{1}(t)$ and $g_{2}(t)$, defined as
\begin{eqnarray}
g_{1}(t)&=&\frac{4}{\vecq^{2}}
\frac{\partial V(\vecq,t)}{\partial t},\label{g1}\\
g_{2}(t)&=&\frac{4}{\vecq^{2}}
\left[V(\vecq,t)+\frac{1}{2}\vecq\;
\frac{\partial V(\vecq,t)}{\partial\vecq}\right],\label{g2}
\end{eqnarray}
equally constitute {\em known\/} functions of time only.
With the coefficients~(\ref{g1}) and (\ref{g2}), the system of
energy and second-moment equations~(\ref{exteq}) may be
reformulated as a {\em linear}, homogeneous third-order system
with time-dependent coefficients,
\begin{equation}\label{auxsys}
\frac{\d}{\d t}\left(\begin{array}{c}\HC\\ -\onehalf\vecq\,\vecp\\
\quarter\vecq^{2}\end{array}\right)=\left(
\begin{array}{ccc}\hphantom{-}0&\hphantom{-}0&g_{1}(t)\\
-1&\hphantom{-}0&g_{2}(t)\\\hphantom{-}0&-1&0\end{array}\right)
\left(\begin{array}{c}\HC\\ -\onehalf\vecq\,\vecp\\
\quarter\vecq^{2}\end{array}\right).
\end{equation}
In conjunction with the full set of canonical equations~(\ref{caneq}),
the system (\ref{auxsys}) does not contain unknown functions---and
hence can be integrated.
In other words, the linear system~(\ref{auxsys}) constitutes an
{\em extension\/} of the system of canonical equations~(\ref{caneq}).

For a further {\em analytic\/} treatment, we will now turn
over to the adjoint system of Eq.~(\ref{auxsys}), i.e., to the
system with the negative transpose system matrix
\begin{equation}\label{adauxsys}
\dot{\bxi}=A(t)\,\bxi,
\qquad\; A(t)=\left(\begin{array}{ccc}0&1&0\\0&0&1\\
-g_{1}(t)&-g_{2}(t)&0\end{array}\right),
\end{equation}
with the column vector $\bxi(t)$ defined by
$\left(\xi(t),\eta(t),\zeta(t)\right)^{T}$.
For this particular system matrix $A(t)$, the linear
system~(\ref{adauxsys}) is obviously equivalent to the linear,
homogeneous, and nonautonomous third-order differential equation
\begin{equation}\label{adauxeq}
\dddot{\xi}+g_{2}(t)\,\dot{\xi}+g_{1}(t)\,\xi=0,
\qquad\eta\equiv\dot{\xi},\quad\zeta\equiv\ddot{\xi}.
\end{equation}
Regardless of the particular form of the system's potential
$V(\vecq,t)$ in the Hamiltonian~(\ref{ham0}), the trace of the
system matrix $A(t)$ from Eq.~(\ref{adauxsys}) is {\em always\/} zero.
Thus the Wronski determinant of any $3\times 3$ solution
matrix $\Xi(t)$ of Eq.~(\ref{adauxsys}) is always constant.
Choosing the unit matrix as the initial condition
($\Xi(0)=\Eins$), the Wronski determinant is then unity.
With our particular system~(\ref{adauxsys}) being equivalent to
Eq.~(\ref{adauxeq}), it is important to realize that a solution
matrix $\Xi(t)$, i.e., a matrix satisfying
\begin{displaymath}
\dot{\Xi}(t)=A(t)\,\Xi(t),
\end{displaymath}
has the form
\begin{equation}\label{wronski}
\Xi(t)=\left(\begin{array}{ccc}
\xi_{1}&\xi_{2}&\xi_{3}\\\dot{\xi}_{1}&\dot{\xi}_{2}&\dot{\xi}_{3}\\
\ddot{\xi}_{1}&\ddot{\xi}_{2}&\ddot{\xi}_{3}\end{array}\right),
\qquad\Xi(0)=\Eins,\qquad\det\Xi(t)\equiv1.
\end{equation}
Thus the lines of $\Xi(t)$ occur as zeroth, first,
and second derivatives of linearly independent functions
$\xi_{k}(t),k=1,2,3$ satisfying Eq.~(\ref{adauxeq}).
With $\Xi(t)$ a solution matrix of the adjoint
system~(\ref{adauxsys}), it is well known~(cf., for example,
Ref.~\cite{adrianova}) that a solution matrix $Z(t)$ of the
energy-second-moment system~(\ref{auxsys}) is then given by
the inverse transpose of the solution matrix $\Xi(t)$,
\begin{equation}\label{reci}
Z(t)=\Xi^{-T}(t)\qquad
\Leftrightarrow\qquad Z^{-1}(t)=\Xi^{T}(t).
\end{equation}
For the solution $\Xi(t)$ of Eq.~(\ref{adauxsys}) with $\Xi(0)=\Eins$,
the mapping of the energy-second-moment vector
$\vecs(t)=(\HC,-\onehalf\vecq\vecp,\quarter\vecq^{2})^{T}$ at $t$
into its initial state $\vecs(0)$ at $t=0$ can thus be written in
terms of the transpose solution matrix $\Xi^{T}(t)$ as
\begin{equation}\label{invgsys}
{\left(\begin{array}{c}\HC\\ -\onehalf\vecq\,\vecp\\
\quarter\vecq^{2}\end{array}\right)}_{t=0}=\left(\begin{array}{ccc}
\xi_{1}&\dot{\xi}_{1}&\ddot{\xi}_{1}\\
\xi_{2}&\dot{\xi}_{2}&\ddot{\xi}_{2}\\
\xi_{3}&\dot{\xi}_{3}&\ddot{\xi}_{3}\end{array}\right)
{\left(\begin{array}{c}\HC\\ -\onehalf\vecq\,\vecp\\
\quarter\vecq^{2}\end{array}\right)}_{t}.
\end{equation}
For the general class of Hamiltonian
systems~(\ref{ham0}), Eq.~(\ref{invgsys}) thus
expresses the fact that the particular vector
$(\HC,-\onehalf\vecq\vecp,\quarter{\vecq^{2}})^{T}$
depends {\em linearly\/} on its initial state at $t=0$, and that
this mapping is associated with a {\em unit determinant}.
We can thus interpret the linear and ``area-preserving''
mapping~(\ref{invgsys}) in the following way:
\begin{quote}
A regular or irregular time evolution of a {\em nonlinear\/}
mapping $(\vecq,\vecp)_{t=0}\mapsto(\vecq,\vecp)_t$
is reflected by the properties of the time evolution
of the ``transfer matrix'' $\Xi^{T}(t)$
that describes the {\em linear\/} reverse mapping
${(\HC,-\onehalf\vecq\vecp,\quarter\vecq^{2})}_{t}
\mapsto {(\HC,-\onehalf\vecq\vecp,\quarter\vecq^{2})}_{t=0}$.
\end{quote}
We emphasize that due to the dependence of the coefficients
$g_{1}(t),g_{2}(t)$ on the spatial trajectory $\vecq(t)$, the
particular solution matrix $\Xi(t)$ of (\ref{adauxsys}) only
applies to the given initial condition $(\vecq(0),\vecp(0))$;
hence we encounter {\em different\/} coefficients
$\bar{g}_{1}(t)$, $\bar{g}_{2}(t)$ and hence a different
solution matrix $\bar{\Xi}(t)$ for a different initial
condition $(\bar{\vecq}(0),\bar{\vecp}(0))$.

With the matrix $\Xi^{T}(t)$ furnishing the correlation
of the ``energy-second-moment vector'' at time $t$ with the vector's
initial state, it is obvious that the properties of the mapping
defined by $\Xi(t)$ reveal information on the system's dynamics.
More precisely, we will show in the following section, that for
any Hamiltonian system~(\ref{ham0}), the Lyapunov analysis of
$\Xi(t)$ yields information on the system's degree of irregularity.
\section{Lyapunov regularity analysis}
\subsection{\label{sec:rgt}Review of the general theory}
If the system matrix $A(t)$ of a nonautonomous
homogeneous linear system
\begin{equation}\label{gensys}
\dot{\bxi}=A(t)\,\bxi
\end{equation}
is nonperiodic, we must resort to Lyapunov's theory of
characteristic exponents (cf., for example,
Refs.~\cite{adrianova,dieci}) to investigate the long term time
behavior of a solution matrix $\Xi(t)$ of Eq.~(\ref{gensys}).
With regard to our particular physical system~(\ref{adauxsys}),
and recalling the definitions of the potential-related terms
$g_{1}(t)$ and $g_{2}(t)$ of Eqs.~(\ref{g1}) and (\ref{g2}),
we may assume the system matrix $A(t)$ from Eq.~(\ref{gensys})
to consist of only bounded coefficients.
The degree of irregularity of the system~(\ref{gensys})
cannot directly be deduced from a solution matrix $\Xi(t)$.
Instead, we must first convert our given system~(\ref{gensys})
into a related system with upper triangular matrix $B(t)$.
According to Perron's theorem on the triangulation of
a (real) linear system, this can always be achieved
by means of an orthogonal transformation:
\begin{adri}(Theorem 3.3.1 of Adrianova\cite{adrianova}).\label{th-tri}
By means of an orthogonal transformation $Q(t)$, any linear
system~(\ref{gensys}) can be reduced to a system with an upper
triangular matrix $B(t)$
\begin{equation}\label{utgensys}
\dot{\bfeta}=B(t)\,\bfeta,\qquad \bxi=Q(t)\,\bfeta.
\end{equation}
\end{adri}
To construct the upper triangular system~(\ref{utgensys}),
we proceed as follows.
A solution matrix $\Xi(t)$ of the system (\ref{gensys})
can always be decomposed into the product of a bounded orthogonal
matrix $Q(t)$ and an upper triangular matrix $R(t)$,
\begin{equation}\label{qrfact}
\Xi(t)=Q(t)\,R(t),\qquad Q^{T}(t)\,Q(t)=\Eins.
\end{equation}
The matrix $Q(t)$ thus defines an orthogonal change of variables
so that the system $\dot{\Xi}(t)=A(t)\,\Xi(t)$ that is defined
by Eq.~(\ref{gensys}) is converted into the triangular system
\begin{displaymath}
\dot{R}(t)=B(t)\,R(t),\qquad B(t)=Q^{T}A(t)Q-Q^{T}\dot{Q}.
\end{displaymath}
The transformation induced by $Q(t)$ is referred to
as a ``Lyapunov transformation.''
Matrices $A(t)$ and $B(t)$ are then called ``kinematicly similar.''
From $\det\Xi(t)\ne0$, we can infer $\det R(t)\ne0$,
as orthogonal matrices $Q(t)$ always have unit determinants.
Therefore the inverse matrix $R^{-1}(t)$ exists, and the
upper triangular matrix $B(t)$ is obtained as
\begin{equation}\label{bmatrix}
B(t)=\dot{R}(t)\,R^{-1}(t).
\end{equation}
Defining the Lyapunov functions $\lambda_{k}(t)$ as the time
averages of the diagonal elements $b_{kk}(t)$ of $B(t)$,
\begin{equation}\label{deflp}
\lambda_{k}(t)=\frac{1}{t}\int_{0}^{t}b_{kk}(\tau)\,\d\tau,
\end{equation}
the Lyapunov characteristic exponents $\lambda^{\text{i}}_{k}$
and $\lambda^{\text{s}}_{k}$ are then given by the limit
values of the $\lambda_{k}(t)$, hence by
\begin{equation}\label{lyapce}
\lambda^{\text{i}}_{k}=\liminf_{t\to\infty}\lambda_{k}(t),\quad
\lambda^{\text{s}}_{k}=\limsup_{t\to\infty}\lambda_{k}(t),\quad k=1,\ldots,n.
\end{equation}
The upper and lower characteristic exponents~(\ref{lyapce})
can now be used to distinguish a regular time evolution of
the solution of Eq.~(\ref{gensys}) from an irregular one.
To this end, we make use of a theorem proven by Lyapunov:
\begin{adri}(Theorem 3.8.1 of Adrianova\cite{adrianova}).\label{th-reg}
A $n\times n$ triangular system $\dot{R}(t)=B(t)\,R(t)$
is regular if and only if the limit values $\lambda^{\text{i}}_{k}$ and
$\lambda^{\text{s}}_{k}$ from Eq.~(\ref{deflp}) coincide, i.e., if
\begin{displaymath}
\lambda^{\text{i}}_{k}=\lambda^{\text{s}}_{k},\qquad k=1,\ldots,n.
\end{displaymath}
\end{adri}
The degree of irregularity of the energy-second-moment map
from Eq.~(\ref{invgsys}) can then directly be inferred from
the degree of irregularity of its adjoint because of the
following fact:
\begin{adri}(Corollary \,3.6.1\, of\, Adrianova\cite{adrianova}).\label{th-adj}
The adjoint system of a (ir)regular system is (ir)regular.
\end{adri}
For a proof of Theorems~\ref{th-tri},~\ref{th-reg},
and~\ref{th-adj} see, Ref.~\cite{adrianova}.
The Lyapunov coefficient of irregularity $\sigma_{\Lambda}$
is then obtained from the limit values~(\ref{lyapce}) as
\begin{equation}\label{lyirrcoeff}
\sigma_{\Lambda}=\sum_{k=1}^{n}\left(\lambda_{k}^{\text{s}}-
\lambda_{k}^{\text{i}}\right).
\end{equation}
If $\lambda_{k}^{\text{i}}=\lambda_{k}^{\text{s}},\;k=1,\ldots,n$, then
Lyapunov's coefficient of irregularity vanishes: $\sigma_{\Lambda}=0$.
In that case, the trajectory is referred to as regular (as defined
by Lyapunov).
Otherwise, we have $\sigma_{\Lambda}>0$, which means that
the system's time evolution exhibits an irregular behavior.
The degree of irregularity of a trajectory of the given dynamic system
can thus be quantified in terms of the value of $\sigma_{\Lambda}\ge0$.
\subsection{\label{sec:gtdh}General (time-dependent) Hamiltonian}
Following the sketched scheme, we will now work out the Lyapunov
functions~(\ref{deflp}) for the particular matrix $\Xi(t)$
from Eq.~(\ref{wronski}) that constitutes a solution of the adjoint
system~(\ref{adauxsys}) of the energy-second-moment system~(\ref{auxsys}).
It turns out that the $QR$ decomposition of Eq.~(\ref{wronski})
and the subsequent time integration of the diagonal elements
$b_{kk}(t), k=1,2,3$ of $B(t)$ from Eq.~(\ref{bmatrix})
yield {\em simple analytical expressions\/} for the
Lyapunov functions~(\ref{deflp}).
With the following abbreviations:
\begin{eqnarray*}
a_{1}(t)&\!\!=\!\!&\xi_{1}^{2}+\dot{\xi}_{1}^{2}+\ddot{\xi}_{1}^{\,2},\\
a_{2}(t)&\!\!=\!\!&{\big(\dot{\xi}_{1}\ddot{\xi}_{2}-
\ddot{\xi}_{1}\dot{\xi}_{2}\big)}^{2}+
{\big(\ddot{\xi}_{1}\xi_{2}-\xi_{1}\ddot{\xi}_{2}\big)}^{2}+
{\big(\xi_{1}\dot{\xi}_{2}-\dot{\xi}_{1}\xi_{2}\big)}^{2},
\end{eqnarray*}
the Lyapunov functions~(\ref{deflp}) for $\Xi(t)$
from Eq.~(\ref{wronski}) are obtained as
\begin{equation}\label{glyapfunct}
\lambda_{1}(t)=\frac{1}{2t}\ln{\frac{a_{2}}{a_{1}}},\quad
\lambda_{2}(t)=\frac{1}{2t}\ln{a_{1}},\quad
\lambda_{3}(t)=-\frac{1}{2t}\ln{a_{2}}.
\end{equation}
A ``Maple''\cite{monagan} worksheet that describes the symbolic
calculation of Eqs.~(\ref{glyapfunct}) starting from the solution
matrix~(\ref{wronski}) of Eq.~(\ref{adauxeq}) via
Eqs.~(\ref{qrfact})--(\ref{deflp}) is listed in the Appendix.
As required, the sum over all $\lambda_{k}(t)$ vanishes
since $\Tr A(t)=\Tr B(t)=0$.
We thus encounter the remarkable and fairly general result:
\begin{quote}
The linear third-order equation (\ref{adauxeq}) always has the
{\em analytical representation\/}~(\ref{glyapfunct})
of its Lyapunov functions (\ref{deflp}).
\end{quote}
According to Theorem~\ref{th-reg}, the regularity analysis
for the energy-second-moment map~(\ref{invgsys}) is carried out
by investigating the asymptotic behavior of the three Lyapunov
functions from Eq.~(\ref{glyapfunct}).
Correspondingly, the system is referred to as {\em regular\/}
if and only if the three limit values
$\lim_{t\to\infty}\lambda_{k}(t),\;k=1,2,3$ exist.

We observe that the system is always regular if the $\xi_{k}(t)$
remain bounded since then all Lyapunov functions~(\ref{glyapfunct})
approach the limit value of zero,
\begin{displaymath}
\xi_{k}(t)\;\text{bounded}\quad\Rightarrow
\quad\text{regular\;trajectory}.
\end{displaymath}
On the other hand, a mere divergence of the $\xi_{k}(t)$ does
{\em not\/} necessarily imply the underlying trajectory be irregular,
\begin{displaymath}
\xi_{k}(t)\;\text{not\;bounded}\quad\not\Rightarrow
\quad\text{irregular\;trajectory}.
\end{displaymath}
An exponential growth of the $\xi_{k}(t)$ that is associated
with {\em sharp\/} asymptotic values of the Lyapunov
functions~(\ref{glyapfunct}) indicates a {\em regular\/}
time evolution of the underlying system trajectory.

The direct numerical calculation of the $\lambda_{k}(t)$ of
Eqs.~(\ref{glyapfunct}) from the solutions of Eq.~(\ref{adauxeq})
is not advisable as the $\xi_{k}(t)$ commonly diverge exponentially.
As a result, very large numerical values of the coefficients $a_{1}(t)$
and $a_{2}(t)$ may occur and hence overflows of floating point numbers.
This {\em numerical\/} problem to calculate $\lambda_{k}(t)$
can be circumvented if we parametrize the functions $a_{1}(t)$
and $a_{2}(t)$ in terms of spherical coordinates.
In order to numerically calculate $\lambda_{2}(t)$,
we define the parametrization~\cite{habib}
\begin{eqnarray}
\cos{\theta}\cos{\psi}&=&\frac{\xi_{1}}{\sqrt{a_{1}}},\quad\;\;\;
\sin{\theta}\cos{\psi}=\frac{\dot{\xi}_{1}}{\sqrt{a_{1}}},\nonumber\\
\sin{\psi}&=&-\frac{\ddot{\xi}_{1}}{\sqrt{a_{1}}},\quad
\theta(0)=0,\quad\psi(0)=0.\label{sepansatz}
\end{eqnarray}
The third-order equation~(\ref{adauxeq}) together with the
expression for $\lambda_{2}$ from Eq.~(\ref{glyapfunct}) are
thus converted into the equivalent nonlinear third-order system
\begin{eqnarray}
\frac{\d}{\d t}\theta(t)&=&-\cos{\theta}\tan{\psi}-
\sin^{2}{\theta},\nonumber\\
\frac{\d}{\d t}\psi(t)&=&\sin{\theta}\sin{\psi}\big(
\sin{\psi}-\cos{\theta}\cos{\psi}\big)\nonumber\\
&&+\cos^{2}{\psi}\big[g_{1}(t)\cos{\theta}+
g_{2}(t)\sin{\theta}\big],\nonumber\\
\frac{\d}{\d t}\big(t\lambda_{2}(t)\big)&=&
\sin{\theta}\cos{\theta}\cos^{2}{\psi}+
\sin{\psi}\cos{\psi}\nonumber\\
&&\times\big[g_{1}(t)\cos{\theta}+
(g_{2}(t)-1)\sin{\theta}\big].
\label{adauxeqa}
\end{eqnarray}
For the numerical calculation of $\lambda_{3}(t)$,
we use the para\-metrization
\begin{eqnarray*}
\cos{\theta}\cos{\psi}&=&\frac{\xi_{1}\dot{\xi}_{2}-
\dot{\xi_{1}}\xi_{2}}{\sqrt{a_{2}(t)}},\quad\;\;
\sin{\theta}\cos{\psi}=\frac{\ddot{\xi}_{1}\xi_{2}-
\xi_{1}\ddot{\xi_{2}}}{\sqrt{a_{2}(t)}},\\
\sin{\psi}&=&-\frac{\dot{\xi}_{1}\ddot{\xi_{2}}-
\ddot{\xi}_{1}\dot{\xi_{2}}}{\sqrt{a_{2}(t)}},\quad
\theta(0)=0,\quad\psi(0)=0.
\end{eqnarray*}
The Lyapunov function $\lambda_{3}(t)$ from Eqs.~(\ref{glyapfunct})
can then be obtained by means of solving the nonlinear system
\begin{eqnarray}
\frac{\d}{\d t}\theta(t)&=&\sin^{2}{\theta}+\cos{\theta}\tan{\psi}+
g_{2}(t)\cos^{2}{\theta},\nonumber\\
\frac{\d}{\d t}\psi(t)&=&-\sin{\theta}\sin^{2}{\psi}-
\cos{\theta}\cos{\psi}\nonumber\\
&&\times\big[g_{1}(t)\cos{\psi}+(g_{2}(t)-1)
\sin{\theta}\sin{\psi}\big],\nonumber\\
\frac{\d}{\d t}\big(t\lambda_{3}(t)\big)&=&
-\sin{\theta}\sin{\psi}\cos{\psi}+\cos{\theta}\cos{\psi}\nonumber\\
&&\!\!\!\times\big[g_{1}(t)\sin{\psi}-(g_{2}(t)-1)
\sin{\theta}\cos{\psi}\big].\label{adauxeqb}
\end{eqnarray}
According to Eqs.~(\ref{glyapfunct}), the sum over the three
Lyapunov functions $\lambda_{1,2,3}(t)$ vanishes.
The remaining function $\lambda_{1}(t)$ is thus given by
$\lambda_{1}(t)=-\lambda_{2}(t)-\lambda_{3}(t)$.
\subsection{\label{sec:atih}Autonomous (time-independent)
Hamiltonian}
If the given system~(\ref{ham0}) is autonomous
(\mbox{$\partial V/\partial t\equiv0$}), then $g_{1}(t)\equiv0$.
Hence $\xi_{1}(t)\equiv1$ is obviously a particular solution
of the linear equation~(\ref{adauxeq}).
With regard to the energy-second-moment map~(\ref{invgsys}),
this solution simply represents the fact that the {\em value\/}
$\HC(t)$ of the Hamiltonian $H$ is a constant of motion
$[\HC(t)=\HC(0)]$ if $H$ does not depend on time explicitly.
For this particular case, we have $a_{1}(t)\equiv1$ and
$a_{2}(t)=\dot{\xi}_{2}^{2}+\ddot{\xi}_{2}^{2}$, so
that the Lyapunov functions~(\ref{glyapfunct}) simplify to
\begin{equation}\label{lyapfunct}
\lambda_{1}(t)=\frac{1}{t}\ln\sqrt{\dot{\xi}_{2}^{2}(t)+
\ddot{\xi}_{2}^{2}(t)},\;\;
\lambda_{2}(t)=0,\;\;
\lambda_{3}(t)=-\lambda_{1}(t).
\end{equation}
Since only the first and second time derivatives of $\xi_{2}(t)$
are needed to calculate $\lambda_{1}(t)$ according to
Eq.~(\ref{lyapfunct}), the Lyapunov functions for autonomous
Hamiltonian systems can be obtained more easily from the
solutions \mbox{$\phi(t)\equiv\dot{\xi}_{2}(t)$} and
\mbox{$\dot{\phi}(t)\equiv\ddot{\xi}_{2}(t)$}
of the Hill-type initial-value problem
\begin{equation}\label{adauxeq1}
\ddot{\phi}+g_{2}(t)\,\phi=0,\qquad\phi(0)=1,\;\dot{\phi}(0)=0,
\end{equation}
with $g_{2}(t)$ the time-dependent coefficient from Eq.~(\ref{g2})
containing a not explicitly time-dependent potential $V(\vecq)$
\begin{displaymath}
g_{2}(t)=\frac{4}{\sum_{i}q_{i}^{2}}\left[V(\vecq)+\frac{1}{2}
\sum_{i=1}^{n}q_{i}\frac{\partial V(\vecq)}{\partial q_{i}}\right].
\end{displaymath}
Lyapunov's coefficient of irregularity
$\sigma_{\Lambda}=2(\lambda_{1}^{\text{s}}-\lambda_{1}^{\text{i}})$
from Eq.~(\ref{lyirrcoeff}) is then obtained from the limit values
$\lambda_{1}^{\text{s}}=\limsup_{t\to\infty}\lambda_{1}(t)$ and
$\lambda_{1}^{\text{i}}=\liminf_{t\to\infty}\lambda_{1}(t)$
of the single function
\begin{equation}\label{lyapfunct1}
\lambda_{1}(t)=\frac{1}{t}\ln\sqrt{\phi^{2}(t)+\dot{\phi}^{2}(t)}.
\end{equation}
We reiterate that the spatial trajectory $\vecq=\vecq(t)$ must
be the {\em known solution\/} of the canonical equations for
the initial value problem~(\ref{adauxeq1}) to be solvable.

The occurrence of large intermediate values that may occur if we
directly numerically calculate $\lambda_{1}(t)$ according
to Eqs.~(\ref{adauxeq1}) and (\ref{lyapfunct1}) can again be avoided
if we parametrize $\phi(t)$ via
\begin{displaymath}
\sin{\psi}=-\frac{\dot{\phi}}{\sqrt{\phi^{2}+\dot{\phi}^{2}}},\;\;
\cos{\psi}=\sqrt{1-\sin^{2}{\phi}}=\frac{\phi}{\sqrt{\phi^{2}+\dot{\phi}^{2}}}.
\end{displaymath}
The linear second-order equation~(\ref{adauxeq1}), together with
Eq.~(\ref{lyapfunct1}), is thus converted into the equivalent
nonlinear second-order system for $t\lambda_{1}(t)$
with the initial condition $\psi(0)=0$,
\begin{eqnarray}
\frac{\d}{\d t}\psi(t)&=&\sin^{2}{\psi}+g_{2}(t)\cos^{2}{\psi},\nonumber\\
\frac{\d}{\d t}\big[t\lambda_{1}(t)\big]&=&
\big[g_{2}(t)-1\big]\sin{\psi}\cos{\psi}\label{lyapfunct1_num}.
\end{eqnarray}
Investigating the Lyapunov functions $\lambda_{k}(t)$ of
the solutions of various Hamiltonian systems, we will
demonstrate in the following example section that the
time evolution of the Lyapunov functions is indeed related
to a regular or irregular time evolution of the respective
dynamical system.
\section{\label{sec:ne}Numerical examples}
\subsection{\label{sec:hhosci}H\'enon-Heiles oscillator}
\begin{figure}[ht]
\centering\includegraphics[height=\linewidth,angle=-90]{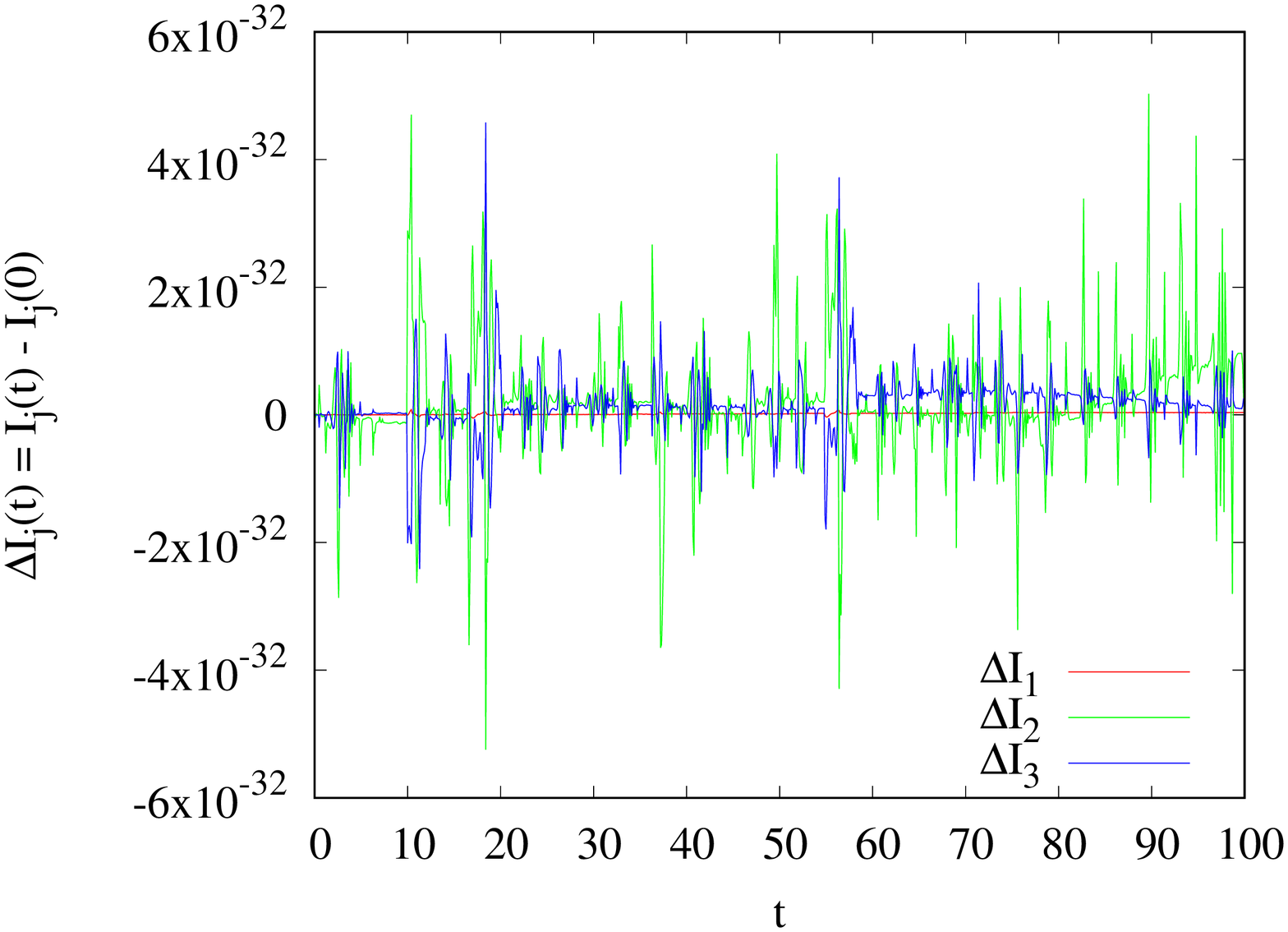}
\caption{Numerical errors $\Delta I_1=h|_{t=0}-h_0$, $\Delta I_2=-\onehalf\vecq\vecp|_{t=0}+\onehalf\vecq_0\vecp_0$,
and $\Delta I_3=\quarter\vecq^{2}|_{t=0}-\quarter\vecq_0^{2}$ of the invariants $I_1=h|_{t=0}$, $I_2=-\onehalf\vecq\vecp|_{t=0}$,
and $I_3=\quarter\vecq^{2}|_{t=0}$, calculated with quadruple precision.}
\label{f0}
\end{figure}

In the first example, we investigate the well-studied
H\'enon-Heiles oscillator\cite{henon,lichtenberg,goldstein,abraham}.
This oscillator models the motion within a two-dimensional
parabolic potential that is perturbed by a cubic potential term.
With the perturbation being proportional to $C$,
its Hamiltonian can be written in normalized form as
\begin{equation}\label{hhdef}
H(\vecq,\vecp)=\onehalf\left(p_{x}^{2}+p_{y}^{2}+
x^{2}+y^{2}\right)+C\left(x^{2}y-\onethird y^{3}\right).
\end{equation}
As the system does not explicitly depend on $t$, we may
restrict ourselves for the regularity analysis to solving the
second-order system for $t\lambda_{1}(t)$ from Eq.~(\ref{lyapfunct1_num}).
The equations of motion and the particular form of the
coefficient $g_{2}(t)$ that characterizes the equation
for $\lambda_{1}(t)$ from Eq.~(\ref{lyapfunct1}) are given by
\begin{eqnarray*}
\ddot{x}+x+2Cxy&=&0\,,\qquad\ddot{y}+y+C(x^{2}-y^{2})=0\\
\ddot{\phi}+g_{2}\,\phi&=&0\,,\qquad
g_{2}=4+10C\,y\frac{x^{2}-y^{2}/3}{x^{2}+y^{2}}.
\end{eqnarray*}
We first verify the energy-second-moment map~(\ref{invgsys}) for the numerical integration of the system~(\ref{hhdef}).
The deviations $\Delta I_j(t)$ of the numerical invariants from the exact invariants given by the initial conditions
$h_0$, $\vecq_0$, and $\vecp_0$ of the numerical integration are plotted in Fig.~\ref{f0}.

\begin{figure}[t]
\begin{minipage}[b]{.49\linewidth}
\centerline{\includegraphics[height=\linewidth,angle=-90]{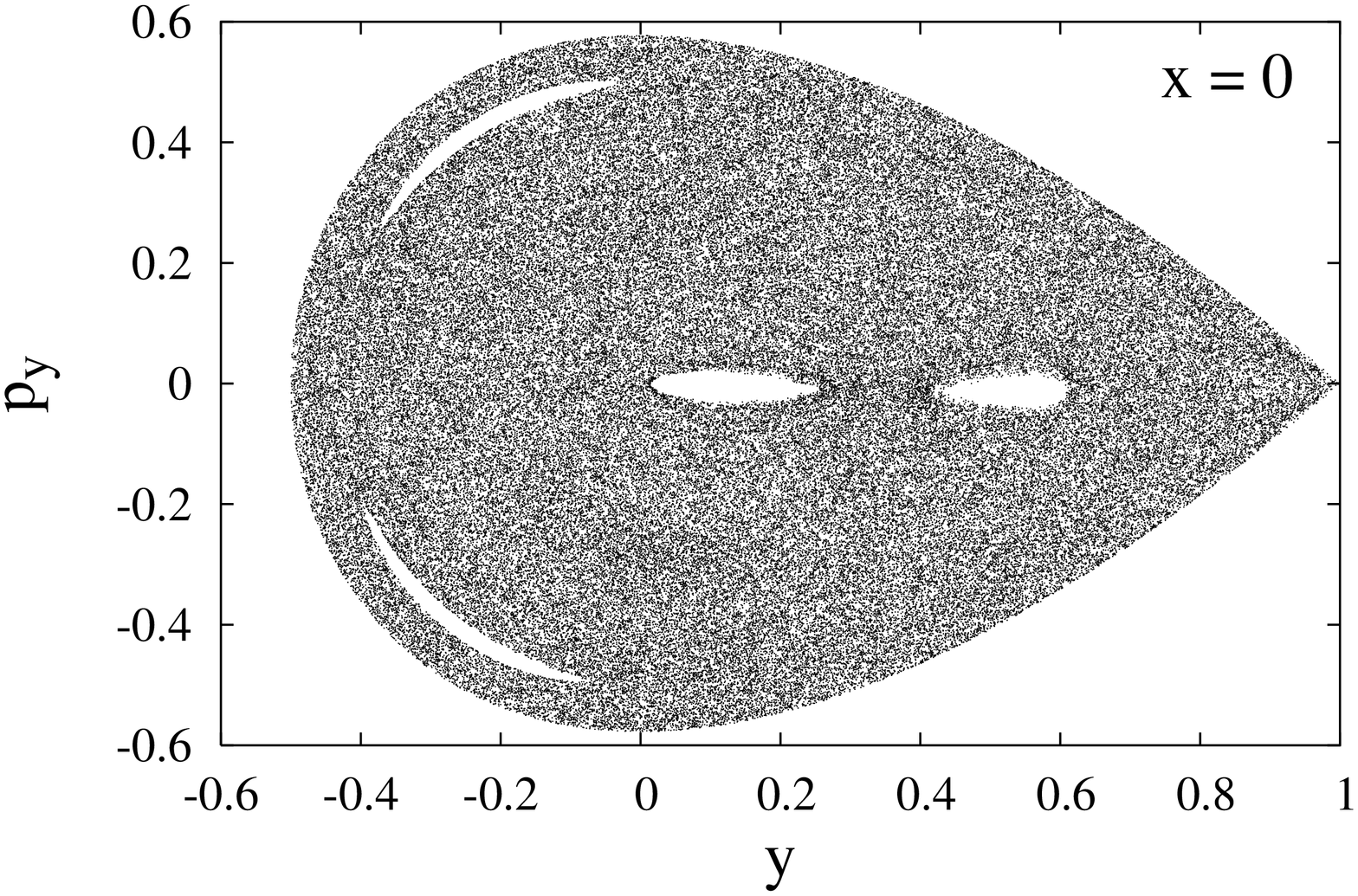}}
\end{minipage}\hfill
\begin{minipage}[b]{.49\linewidth}
\centerline{\includegraphics[height=\linewidth,angle=-90]{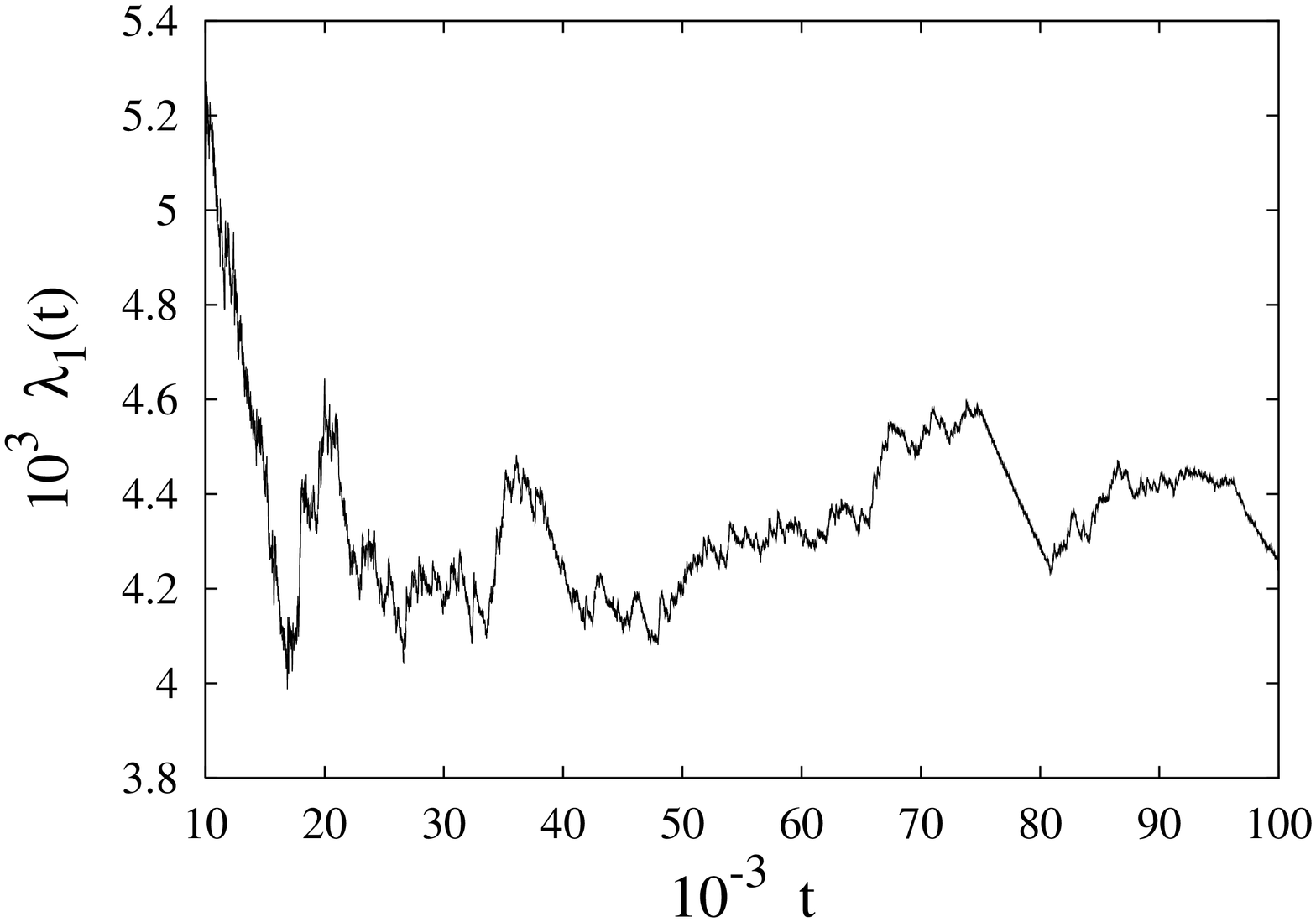}}
\end{minipage}
\caption{\label{f1}Left: $y,p_{y}$-Poincar\'e surface-of-section
representation of an irregular trajectory in the H\'enon-Heiles
oscillator~(\ref{hhdef}) with the limiting energy $\HC=1/6$
for the initial condition
$(x_{0},p_{x,0},y_{0},p_{y,0})=(0,0.5367,-0.2,0)$ and $C=1$.
Right: Lyapunov function $\lambda_{1}(t)$ from
Eq.~(\ref{lyapfunct1_num}) for this trajectory.}
\end{figure}

With the potential of this system being of third order, the
singularity of $g_{2}(t)$ at $(x(t),y(t))=(0,0)$ is {\em removable}.
Setting $C=1$ and fixing the system's dimensionless energy to the
limiting value of $\HC=1/6$ for a bounded motion, we obtain for the
initial condition $(x_{0},p_{x,0},y_{0},p_{y,0})=(0,0.5367,-0.2,0)$
the Poincar\'e surface-of-section displayed on the left-hand-side
of Fig.~\ref{f1}.
The points in this figure display the $(y,p_{y})$-coordinates
of the particle in the course of its time evolution whenever its
$x$-coordinate satisfies the condition $x=0$.
The picture shows that almost the entire available phase-space
area is covered in the course of the trajectory's time
evolution---which indicates that this particular
trajectory is {\em irregular}.
As each of these points can be considered by itself as an initial
condition, we conclude that the vast majority of initial conditions that
are associated with the energy $h=1/6$ give rise to irregular orbits.
\begin{figure}[t]
\begin{minipage}[b]{.49\linewidth}
\centerline{\includegraphics[height=1.1\linewidth,angle=-90]{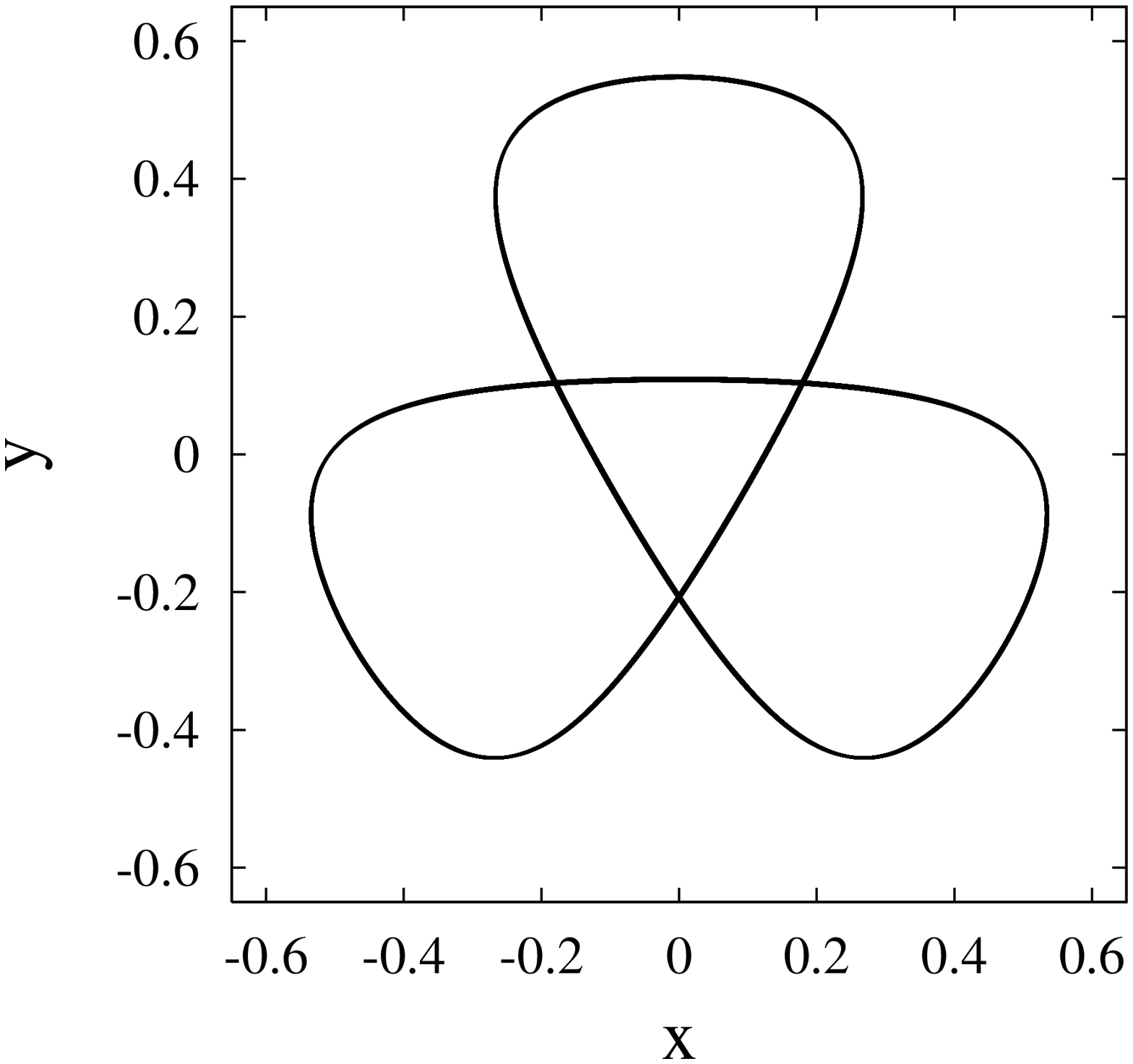}}
\end{minipage}\hfill
\begin{minipage}[b]{.49\linewidth}\vfill
\centerline{\includegraphics[height=\linewidth,angle=-90]{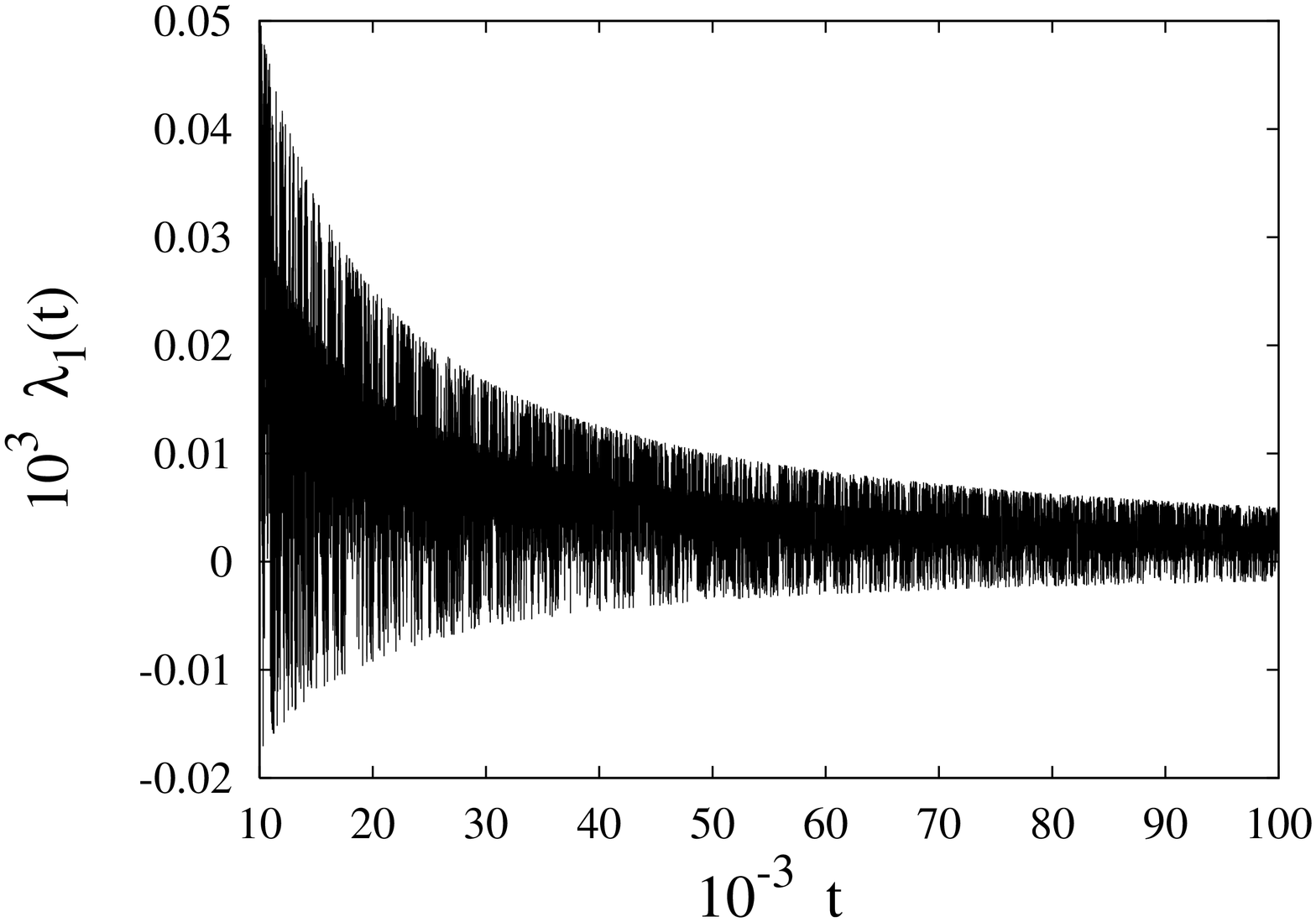}}
\vfill\end{minipage}
\caption{\label{f2}Left: real-space projection of a regular
trajectory of the H\'enon-Heiles oscillator~(\ref{hhdef}) with the
limiting energy $\HC=1/6$ as obtained for the initial condition
$(x_{0},p_{x,0},y_{0},p_{y,0})=(0,0.3765,0.55,0)$ and $C=1$.
Right: Lyapunov function $\lambda_{1}(t)$ from
Eq.~(\ref{lyapfunct1_num}) for this trajectory.}
\end{figure}

\begin{figure}[t]
\begin{minipage}[b]{.49\linewidth}
\centerline{\includegraphics[height=\linewidth,angle=-90]{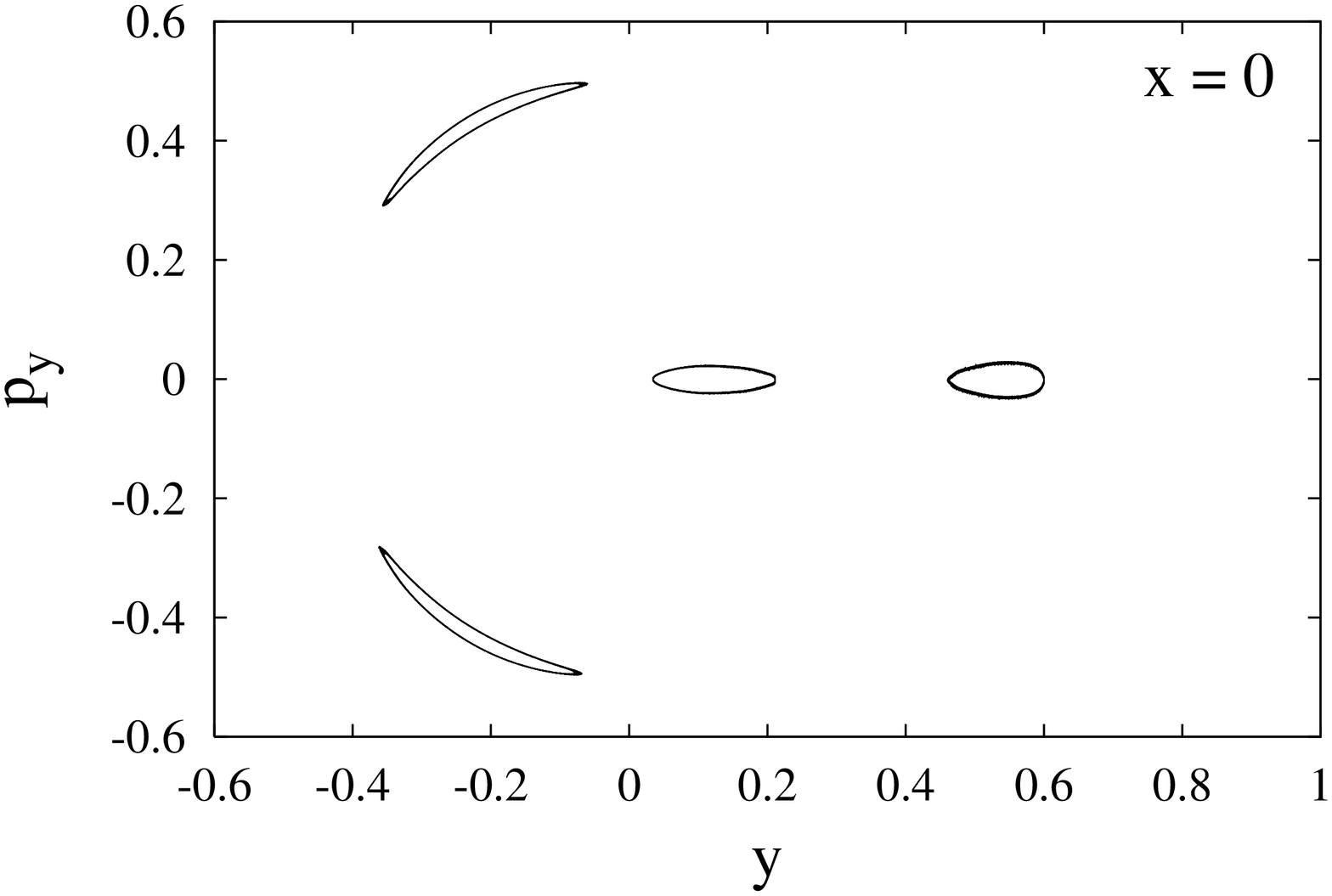}}
\end{minipage}\hfill
\begin{minipage}[b]{.49\linewidth}\vfill
\centerline{\includegraphics[height=\linewidth,angle=-90]{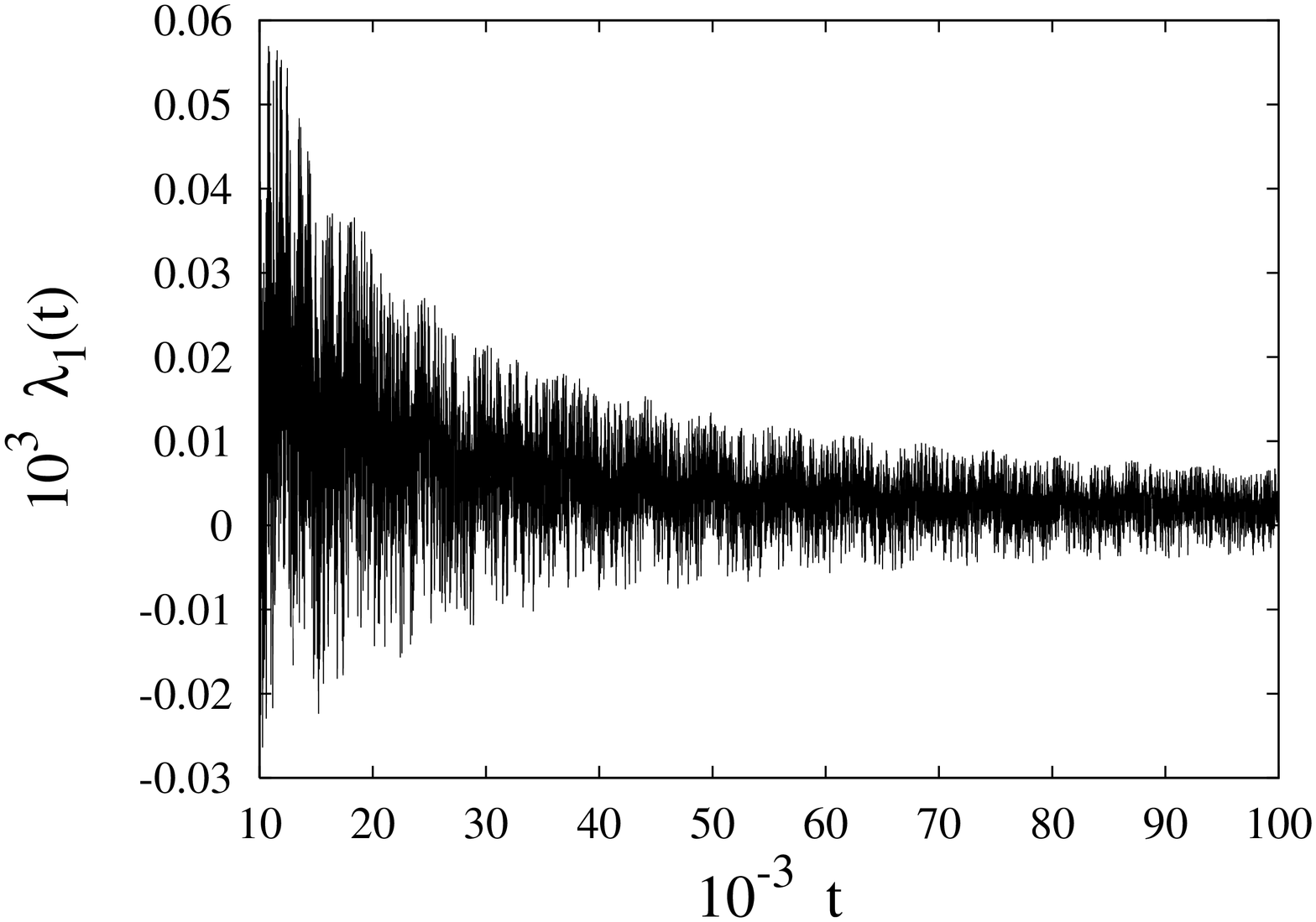}}
\vfill\end{minipage}
\caption{\label{f2a}Left: $y,p_{y}$-Poincar\'e surface-of-section
representation of a regular trajectory near the boundary of the
regular region with the limiting energy $\HC=1/6$ as obtained
for the initial condition
$(x_{0},p_{x,0},y_{0},p_{y,0})=(0,0.3420,0.60,0.02)$ and $C=1$.\newline
Right: Lyapunov function $\lambda_{1}(t)$ from
Eq.~(\ref{lyapfunct1_num}) for this trajectory.}
\end{figure}

In contrast, the blank areas within the dotted region correspond to
the disjunct volume of phase space where the motion is {\em regular}.
A regular trajectory is displayed in the picture on the left
of Fig.~\ref{f2} in the form of a real space projection of the
phase-space motion.
It is obtained by choosing the particle's initial conditions
to lie within the blank islands of the Poincar\'e
surface-of-section of Fig.~\ref{f1}.
We observe in the picture on the left of Fig.~\ref{f2} that the
particle crosses the vertical line $x=0$ exactly four times in
the course of one oscillation period at different slopes $p_{y}$.
These four locations correspond to the four blank islands
occurring in the Poincar\'e section of Fig.~\ref{f1}.
The occurrence of islands of {\em finite area\/} in the
phase-space plane of Fig.~\ref{f1} indicates that a regular
motion is not only given for the strictly closed ``clover-leaf''
motion but also for an oscillation of some finite amplitude
around this closed trajectory.
A Poincar\'e surface-of-section of such a trajectory, together with
its Lyapunov function $\lambda_{1}(t)$, is displayed in Fig.~\ref{f2a}.
\begin{figure}[t]
\begin{minipage}[b]{.49\linewidth}
\centerline{\includegraphics[height=\linewidth,angle=-90]{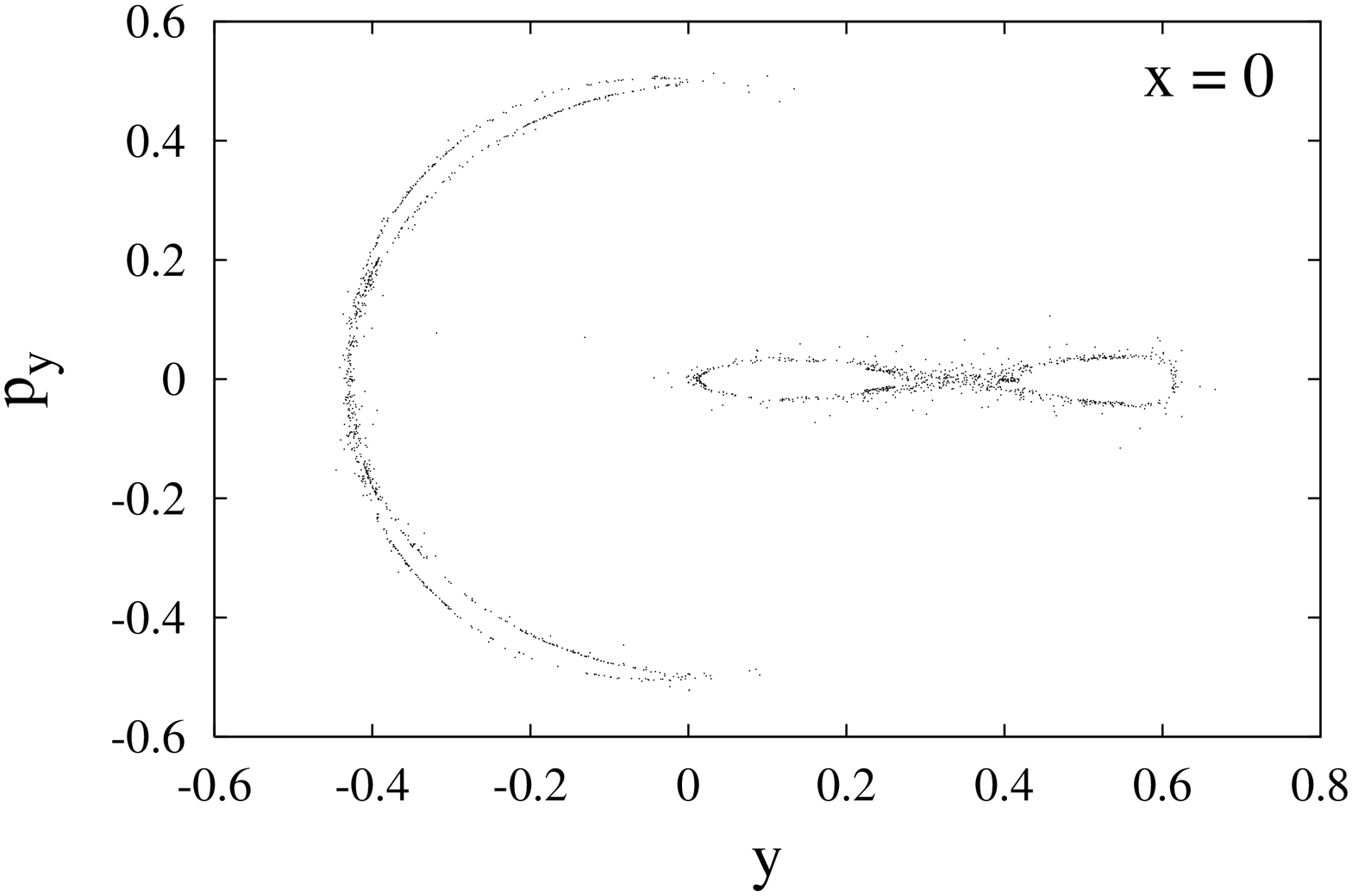}}
\end{minipage}\hfill
\begin{minipage}[b]{.49\linewidth}
\centerline{\includegraphics[height=\linewidth,angle=-90]{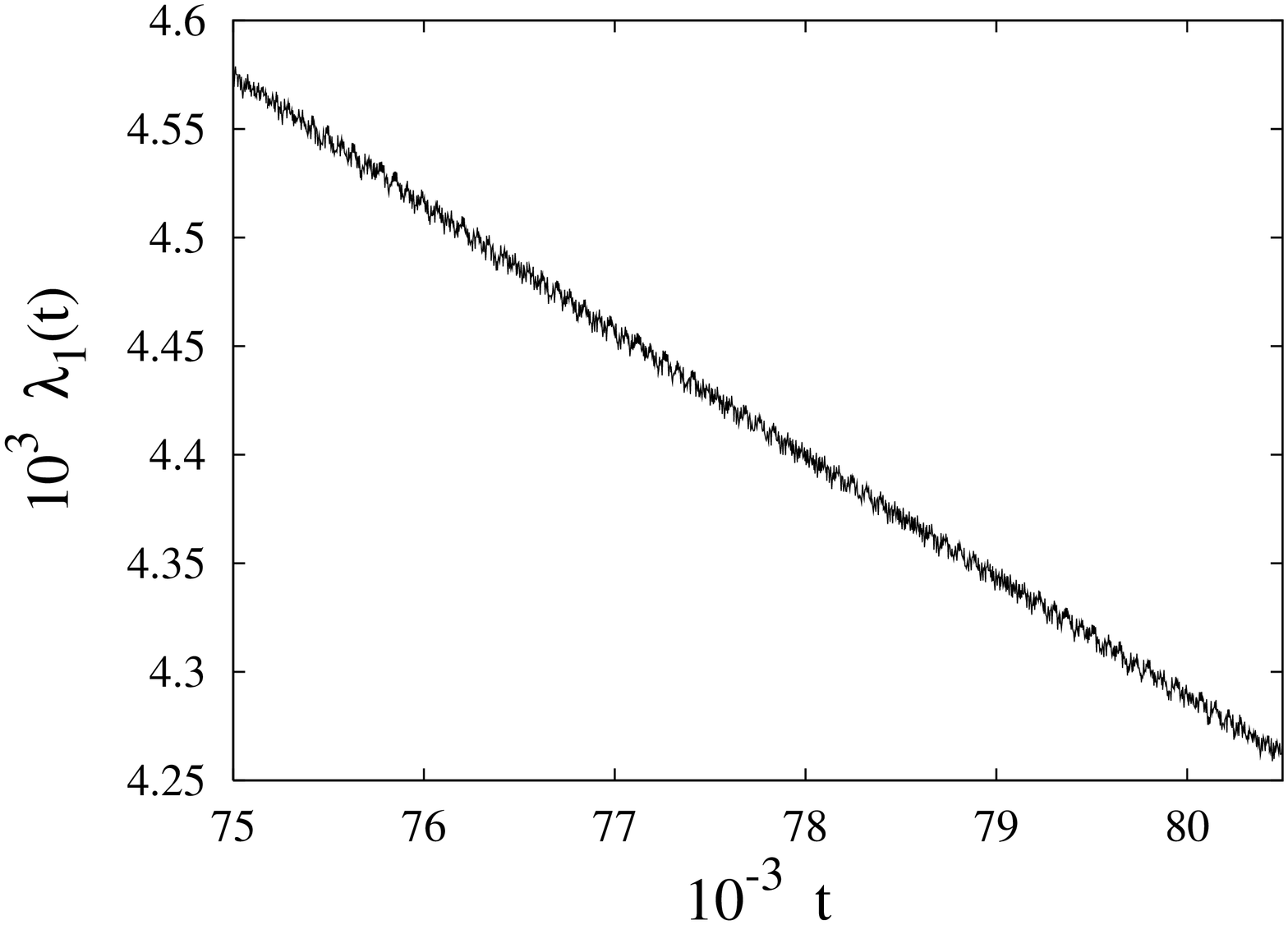}}
\end{minipage}
\caption{\label{f1a}
Right: enlarged view of the Lyapunov function
$\lambda_{1}(t)$ from Fig.~\ref{f1} for the time
interval $75\times 10^{3}\le t\le 80.5\times 10^{3}$.
Left: subset of the $y,p_{y}$-Poincar\'e-section
points of Fig.~\ref{f1} that emerges during that
particular time interval.}
\end{figure}
\begin{figure}[ht]
\begin{minipage}[b]{.49\linewidth}
\centerline{\includegraphics[height=\linewidth,angle=-90]{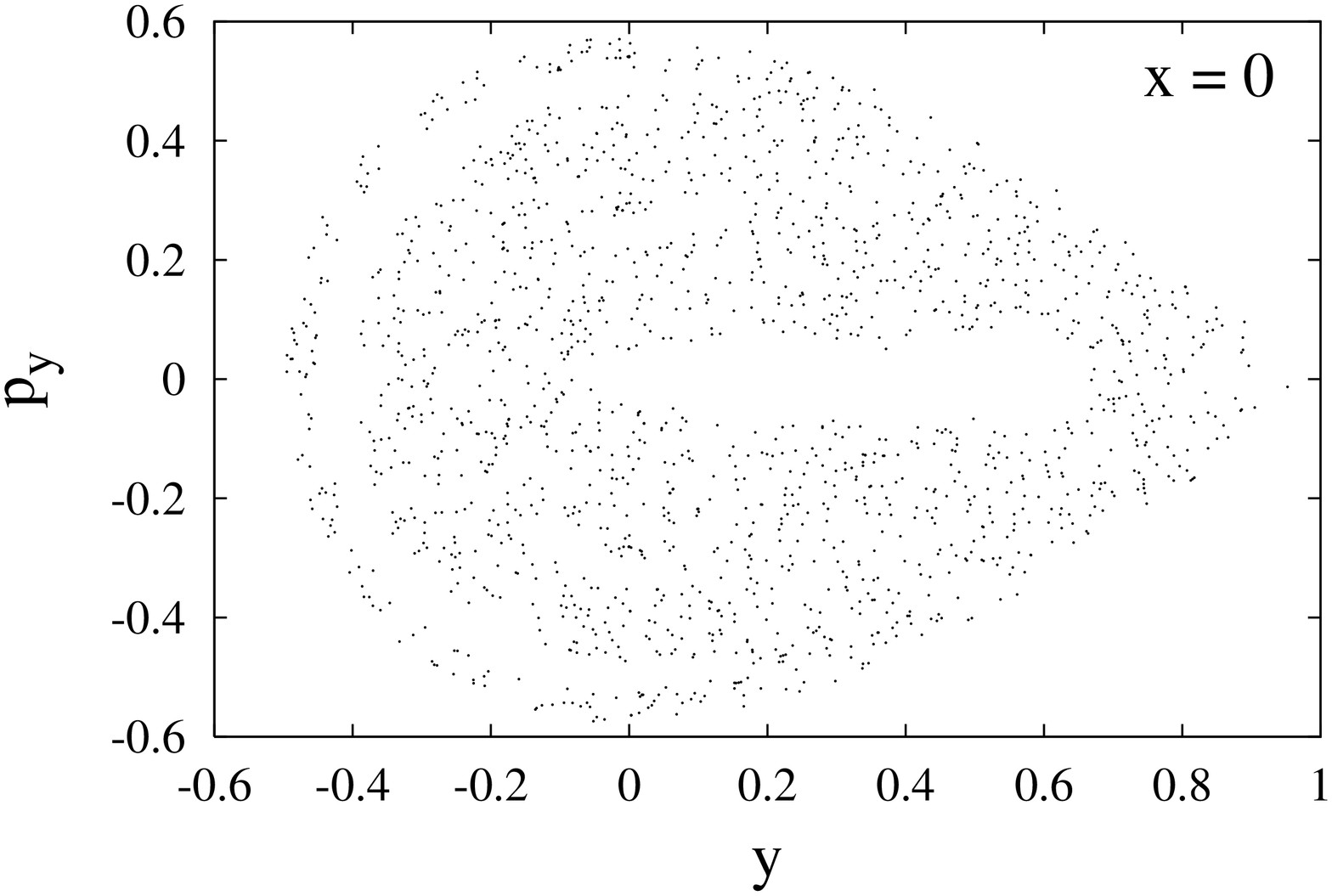}}
\end{minipage}\hfill
\begin{minipage}[b]{.49\linewidth}
\centerline{\includegraphics[height=\linewidth,angle=-90]{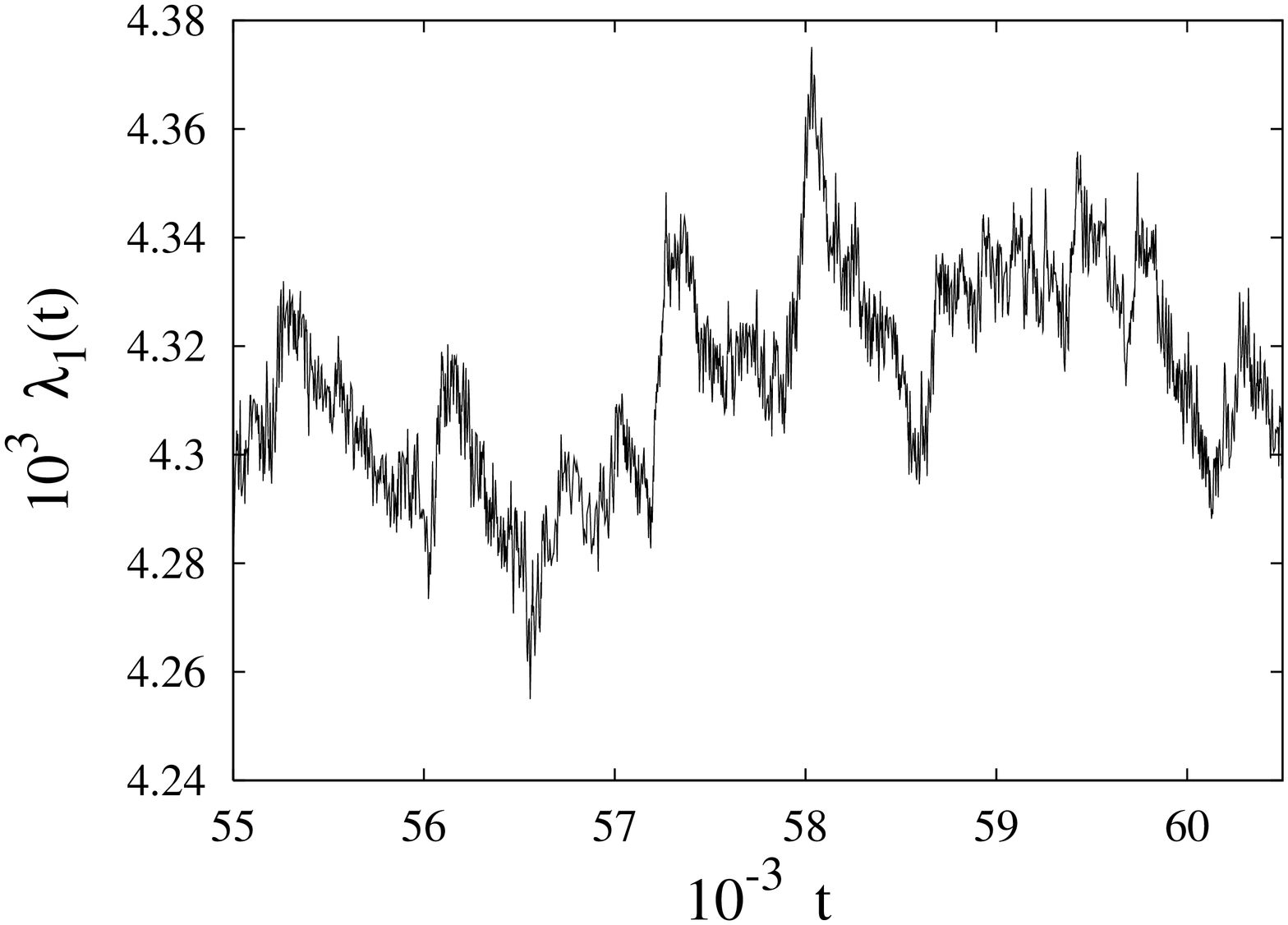}}
\end{minipage}
\caption{\label{f1b}
Right: enlarged view of the Lyapunov function
$\lambda_{1}(t)$ from Fig.~\ref{f1} for the time
interval $55\times 10^{3}\le t\le 60.5\times 10^{3}$.
Left: subset of the $y,p_{y}$-Poincar\'e-section
points of Fig.~\ref{f1} that emerges during that
particular time interval.}
\end{figure}

According to Theorem~\ref{th-reg}, the existence of a sharp
asymptotic value of $\lambda_{1}(t)$ indicates the regularity of
the respective trajectory $(x(t),y(t))$, i.e.,
\begin{equation}\label{lyaplimit}
\lambda_{1}^{i}=\lambda_{1}^{s}=\lim_{t\to\infty}\lambda_{1}(t)
\Longleftrightarrow\text{regular\;trajectory}.
\end{equation}
The solutions $\lambda_{1}(t)$ of Eq.~(\ref{lyapfunct1_num})
for the regular particle motions $(x(t),y(t))$ are displayed
in the pictures on the right of Figs.~\ref{f2} and~\ref{f2a}.
For these particular cases, the solution functions $\lambda_{1}(t)$
approach zero, hence converge to limit values.
This is the expected result for regular trajectories.

In contrast, we observe in the picture on the right of
Fig.~\ref{f1} that the function $\lambda_{1}(t)$ associated
with the irregular particle motion does {\em not\/} converge
to a limit value.
From Theorem~\ref{th-reg}, we conclude that the underlying
trajectory is irregular---which is in agreement with the
impression obtained from the Poincar\'e section of Fig.~\ref{f1}.

Furthermore, we can show for a chaotic trajectory that the
detailed shape of the Lyapunov functions in the course of
the system's time evolution allows us to identify time
intervals where the motion is temporarily quasiregular.
In Fig.~\ref{f1a}, we display an excerpt of Fig.~\ref{f1}
for the time interval $75\times 10^{3}\le t\le 80.5\times 10^{3}$.
The right-hand-side plot of Fig.~\ref{f1a} shows that---apart
from local fluctuations---the $\lambda_{1}$ curve
decreases monotonicly in that time interval.
This behavior of $\lambda_{1}$ resembles the regular cases
displayed in Figs.~\ref{f2} and~\ref{f2a}.
The section points occurring during this particular time
interval are shown in the left-hand side of Fig.~\ref{f1a}.
We observe that these points are {\em not\/} spread over the
possible phase-space region as one would expect for a chaotic
motion with regard to the Poincar\'e section of Fig.~\ref{f1}.
Instead, all section points are located in the vicinity
of the four blank islands of Fig.~\ref{f1}.
This arrangement of section points corresponds to a
``clover-leaf''-like oscillation that is displayed in the
left-hand-side picture of Fig.~\ref{f2}.
We thus find that the chaotic trajectory is temporarily
``trapped'' into a quasiregular behavior.

A converse case is displayed in Fig.~\ref{f1b}.
The right-hand-side plot shows another enlarged portion of
the graph of $\lambda_{1}(t)$ from Fig.~\ref{f1}, namely the
time interval $55\times 10^{3}\le t\le 60.5\times 10^{3}$.
In contrast to the behavior emerging in the case of Fig.~\ref{f1a},
$\lambda_{1}$ keeps on fluctuating rapidly in that particular
time interval and does not converge at times.
The corresponding Poincar\'e-section points are shown
on the left-hand side of Fig.~\ref{f1b}.
We observe that the section points are now spread
over the entire available phase space.
Thus, in this time interval, the system exhibits a purely
chaotic behavior.
Summarizing, we conclude that the detailed shape of the graph
of $\lambda_{1}(t)$ provides us with information on the
``instantaneous chaoticity'' of the underlying trajectory.
\subsection{Nonlinear two-dimensional oscillator}
We now investigate the chaos-order transitions of a model oscillator
consisting of a two-dimensional harmonic oscillator
that is disturbed by a fourth-order potential term depending
on a coupling coefficient $C$.
The characteristics of the trajectory as a function of the
coupling coefficient $C$ in this two-dimensional Hamiltonian
system was studied by Deng and Hioe\cite{deng}.
Its Hamiltonian is given in normalized form by
\begin{equation}\label{dodef}
H(\vecq,\vecp)=\onehalf(p_{x}^{2}+p_{y}^{2}+
x^{2}+y^{2})+\mu(x^{4}+2Cx^{2}y^{2}+y^{4}).
\end{equation}
The equations of motion following from Eq.~(\ref{dodef}) are
\begin{equation}\label{hheqm}
\ddot{x}+x+4\mu\left(x^{3}+Cxy^{2}\right)=0,\quad
\ddot{y}+y+4\mu\left(y^{3}+Cx^{2}y\right)=0.
\end{equation}
As the system~(\ref{dodef}) is autonomous, the coefficient
$g_{1}(t)$ vanishes identically.
For the Lyapunov stability analysis of the energy-second-moment
map~(\ref{invgsys}), we may therefore again restrict
ourselves to solving the second-order system~(\ref{lyapfunct1_num})
and analyzing the single Lyapunov function $\lambda_{1}(t)$.
For the particular potential of Eq.~(\ref{dodef}), the
coefficient $g_{2}(t)$ is given by
\begin{equation}\label{hhadauxeq}
g_{2}(t)=4+12\mu\,\frac{x^{4}(t)+2Cx^{2}(t)\,y^{2}(t)+y^{4}(t)}
{x^{2}(t)+y^{2}(t)}.
\end{equation}
Again, the point $(x(t),y(t))=(0,0)$ is a {\em removable\/}
singularity of $g_{2}(t)$.
In our numerical calculations, we used a fixed scaling
parameter $\mu=1$.
We compared the time evolution of the Lyapunov functions
$\lambda_{1}(t)$ for different values of the coupling constant~$C$.
Following Deng and Hioe~\cite{deng}, we defined for the trajectory's
initial condition $x_{0}=5$, $y_{0}=10$, $p_{x,0}=p_{y,0}=0$
in all calculations.
\begin{figure}[t]
\begin{minipage}[b]{.49\linewidth}
\centerline{\includegraphics[height=\linewidth,angle=-90]{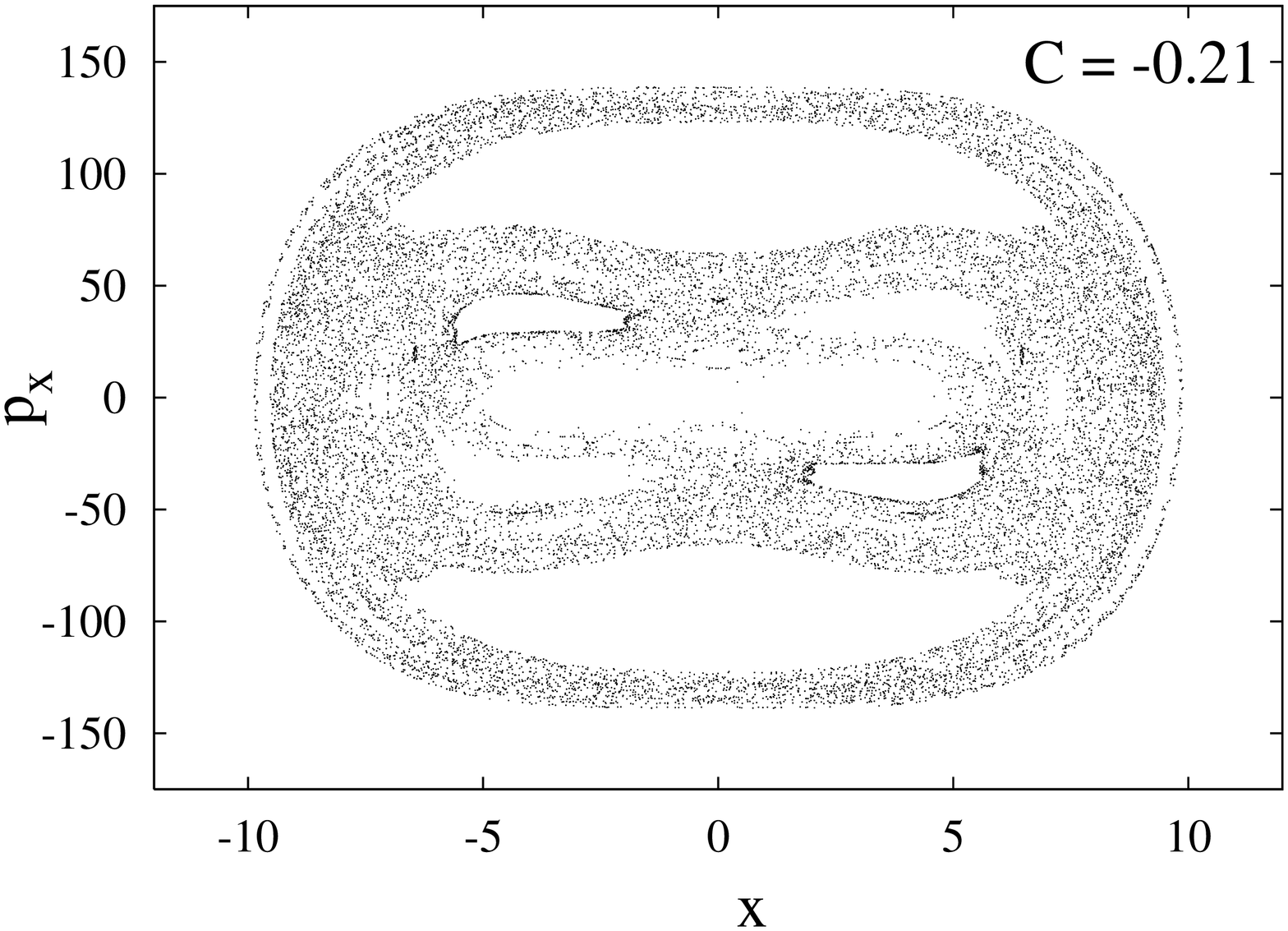}}
\end{minipage}\hfill
\begin{minipage}[b]{.49\linewidth}
\centerline{\includegraphics[height=\linewidth,angle=-90]{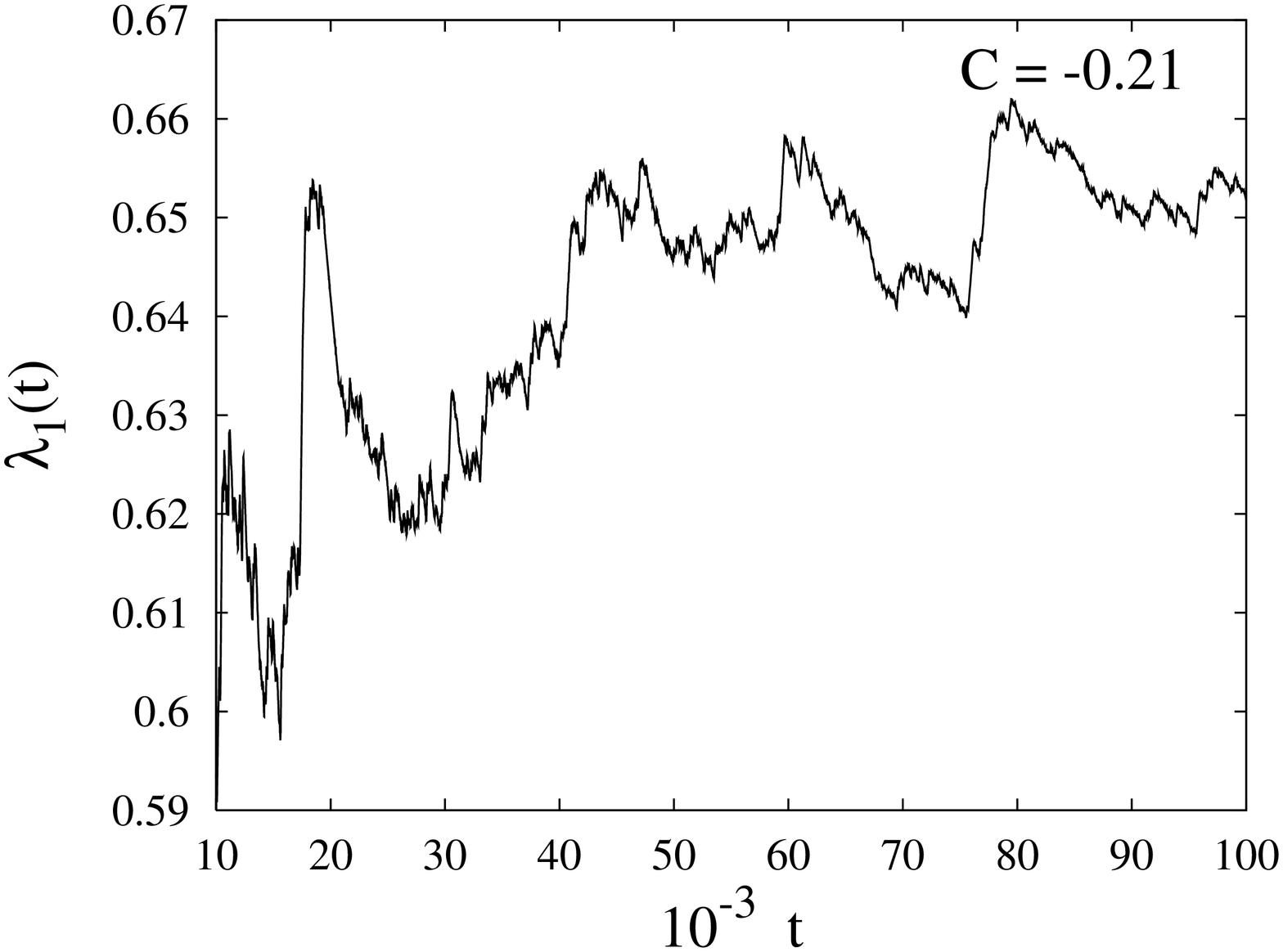}}
\end{minipage}
\begin{minipage}[b]{.49\linewidth}
\centerline{\includegraphics[height=\linewidth,angle=-90]{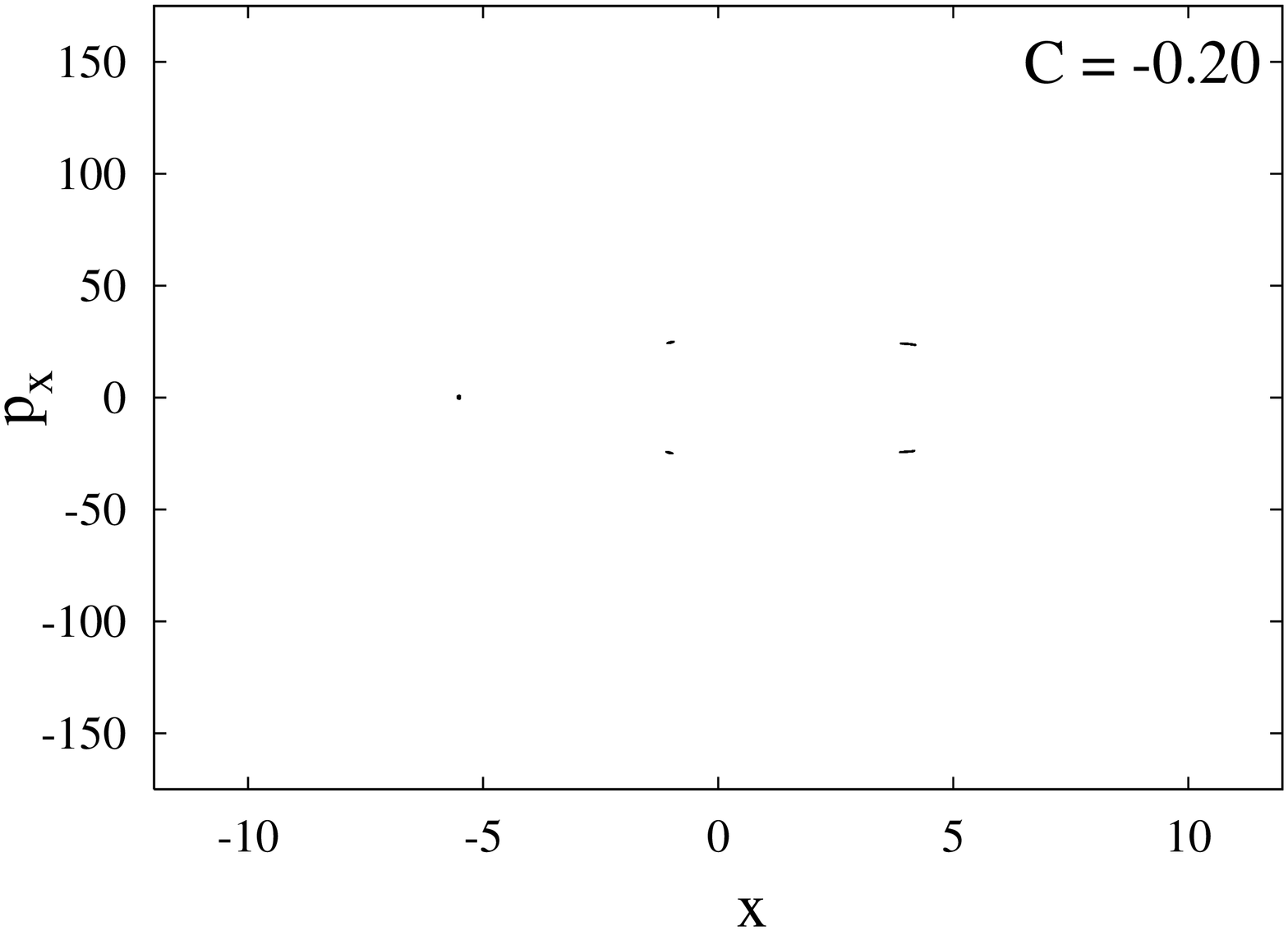}}
\end{minipage}\hfill
\begin{minipage}[b]{.49\linewidth}
\centerline{\includegraphics[height=\linewidth,angle=-90]{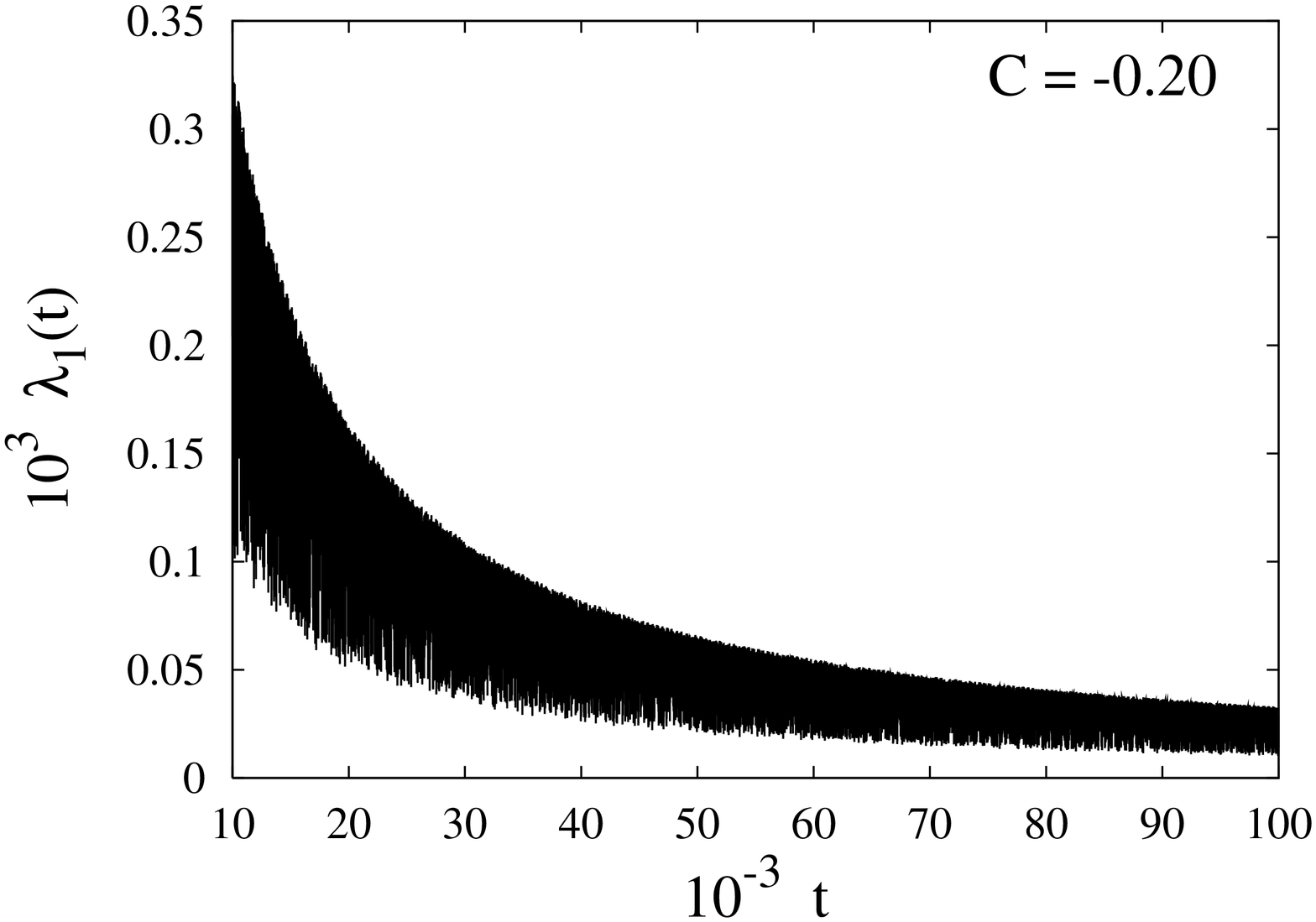}}
\end{minipage}
\begin{minipage}[b]{.49\linewidth}
\centerline{\includegraphics[height=\linewidth,angle=-90]{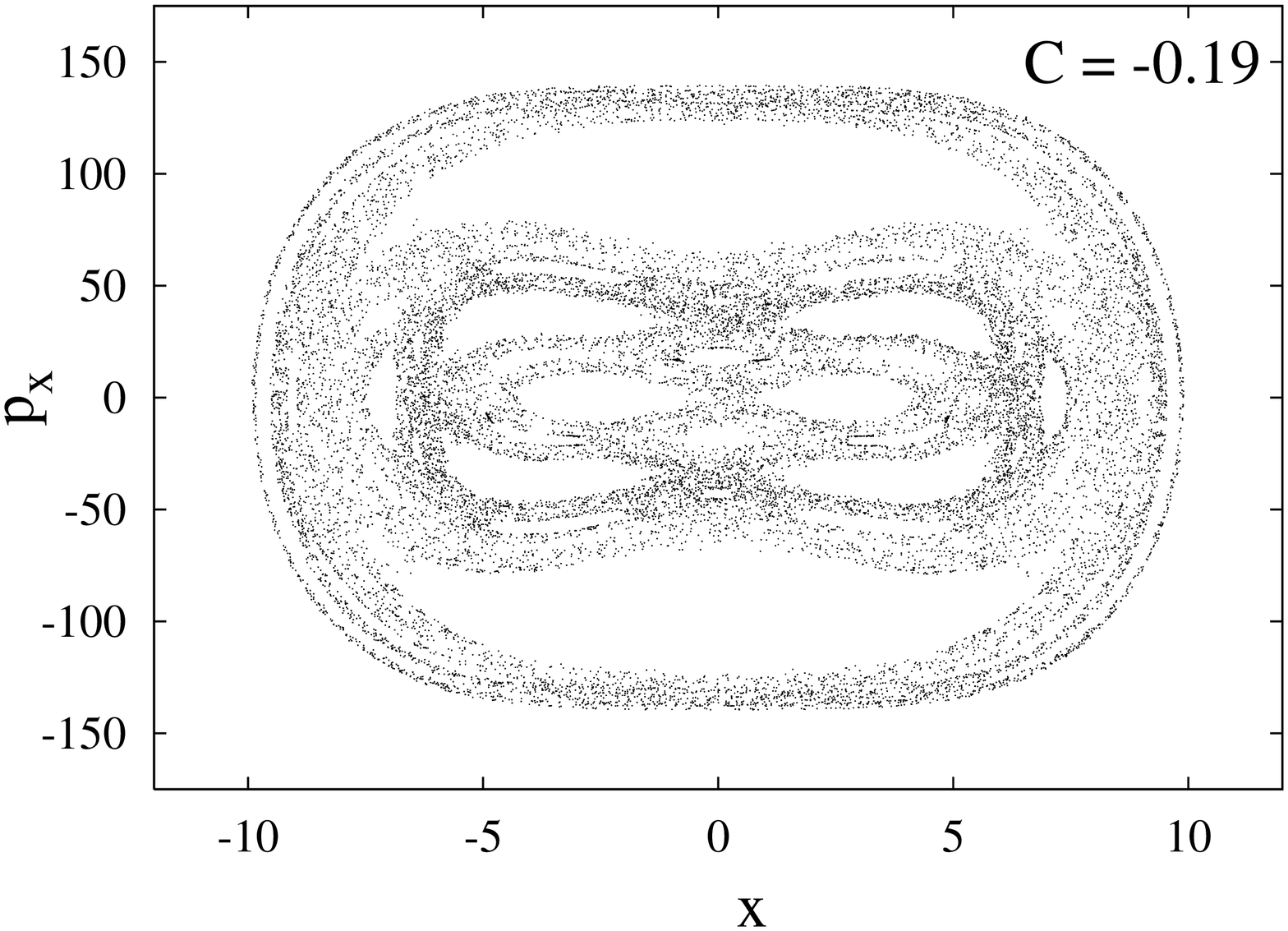}}
\end{minipage}\hfill
\begin{minipage}[b]{.49\linewidth}
\centerline{\includegraphics[height=\linewidth,angle=-90]{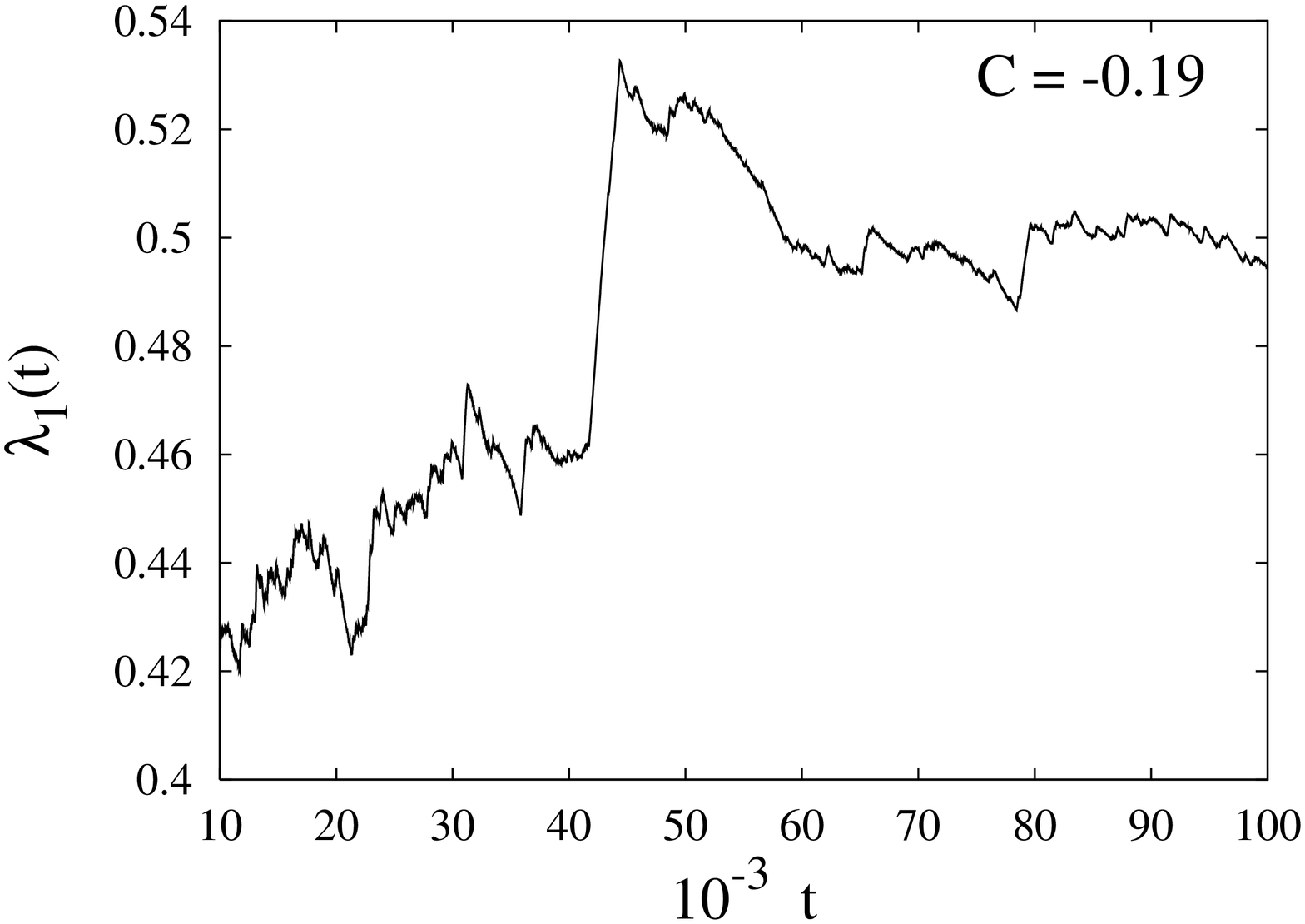}}
\end{minipage}
\caption{Poincar\'e surfaces-of-section (left) and Lyapunov functions
$\lambda_{1}(t)$ from Eq.~(\ref{lyapfunct1_num})
for different values of the coupling constant $C$ of the nonlinear
2D oscillator~(\ref{dodef}).}
\end{figure}
\setcounter{figure}{5}

\begin{figure}[ht]
\begin{minipage}[b]{.49\linewidth}
\centerline{\includegraphics[height=\linewidth,angle=-90]{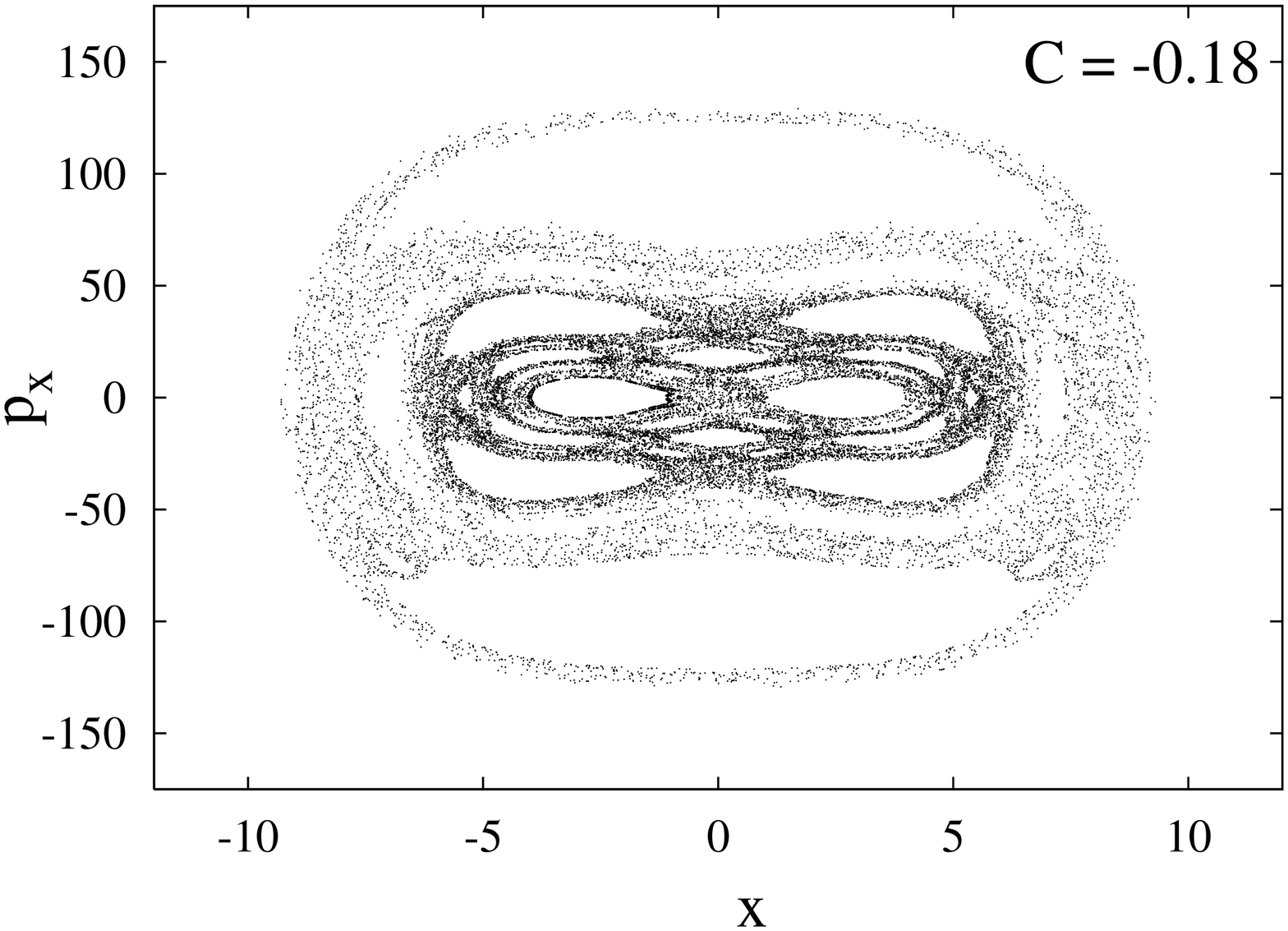}}
\end{minipage}\hfill
\begin{minipage}[b]{.49\linewidth}
\centerline{\includegraphics[height=\linewidth,angle=-90]{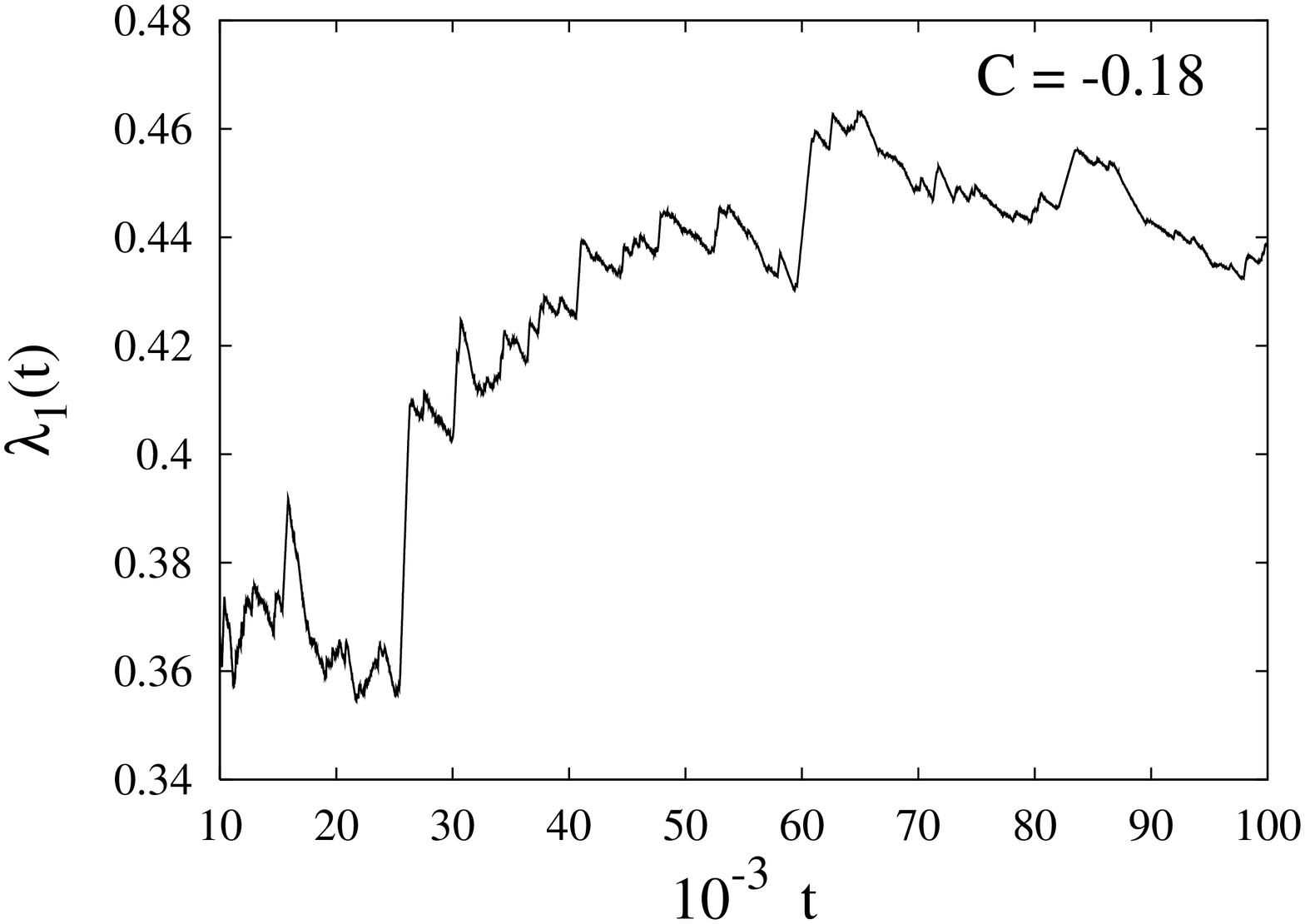}}
\end{minipage}
\begin{minipage}[b]{.49\linewidth}
\centerline{\includegraphics[height=\linewidth,angle=-90]{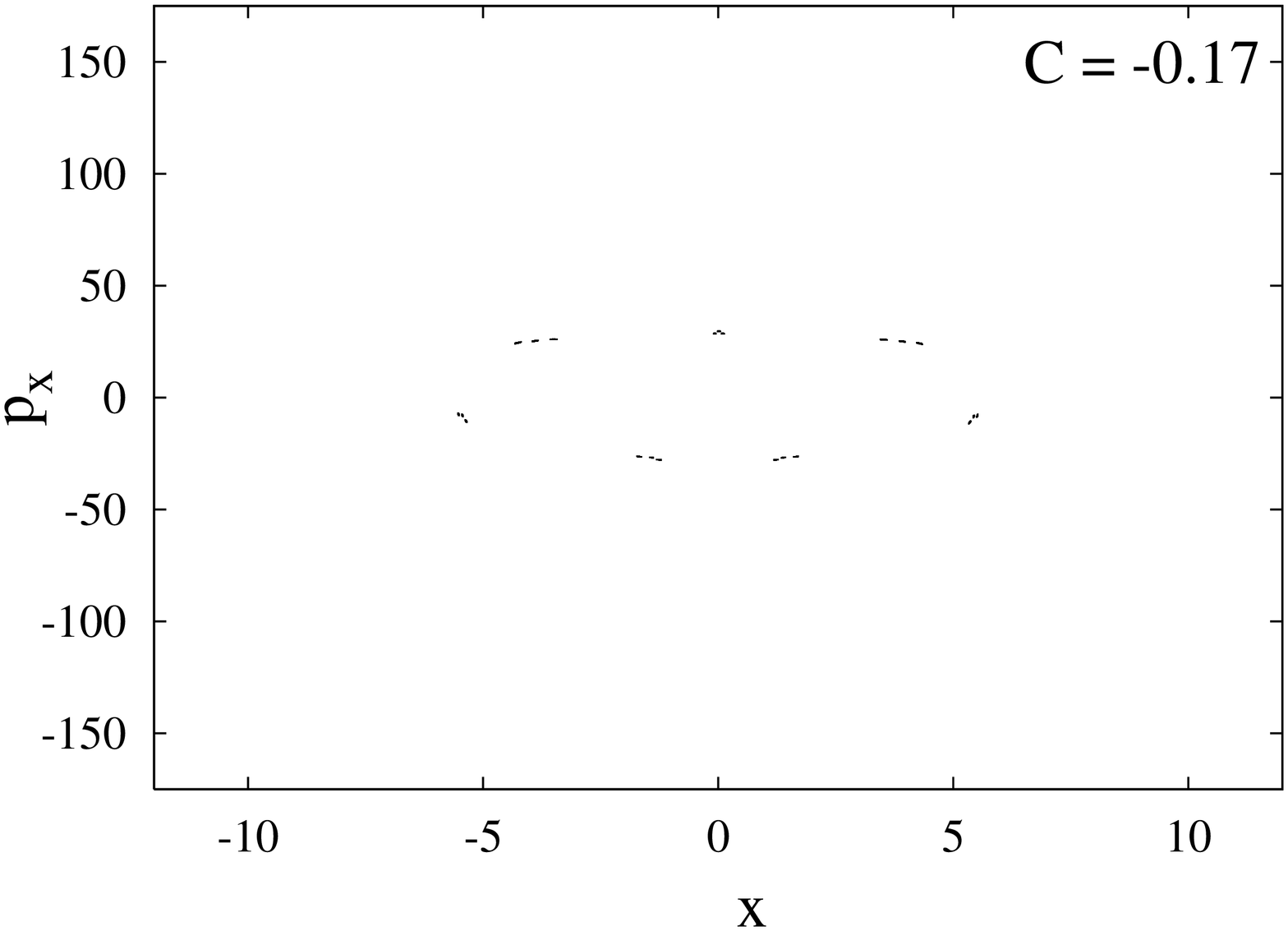}}
\end{minipage}\hfill
\begin{minipage}[b]{.49\linewidth}
\centerline{\includegraphics[height=\linewidth,angle=-90]{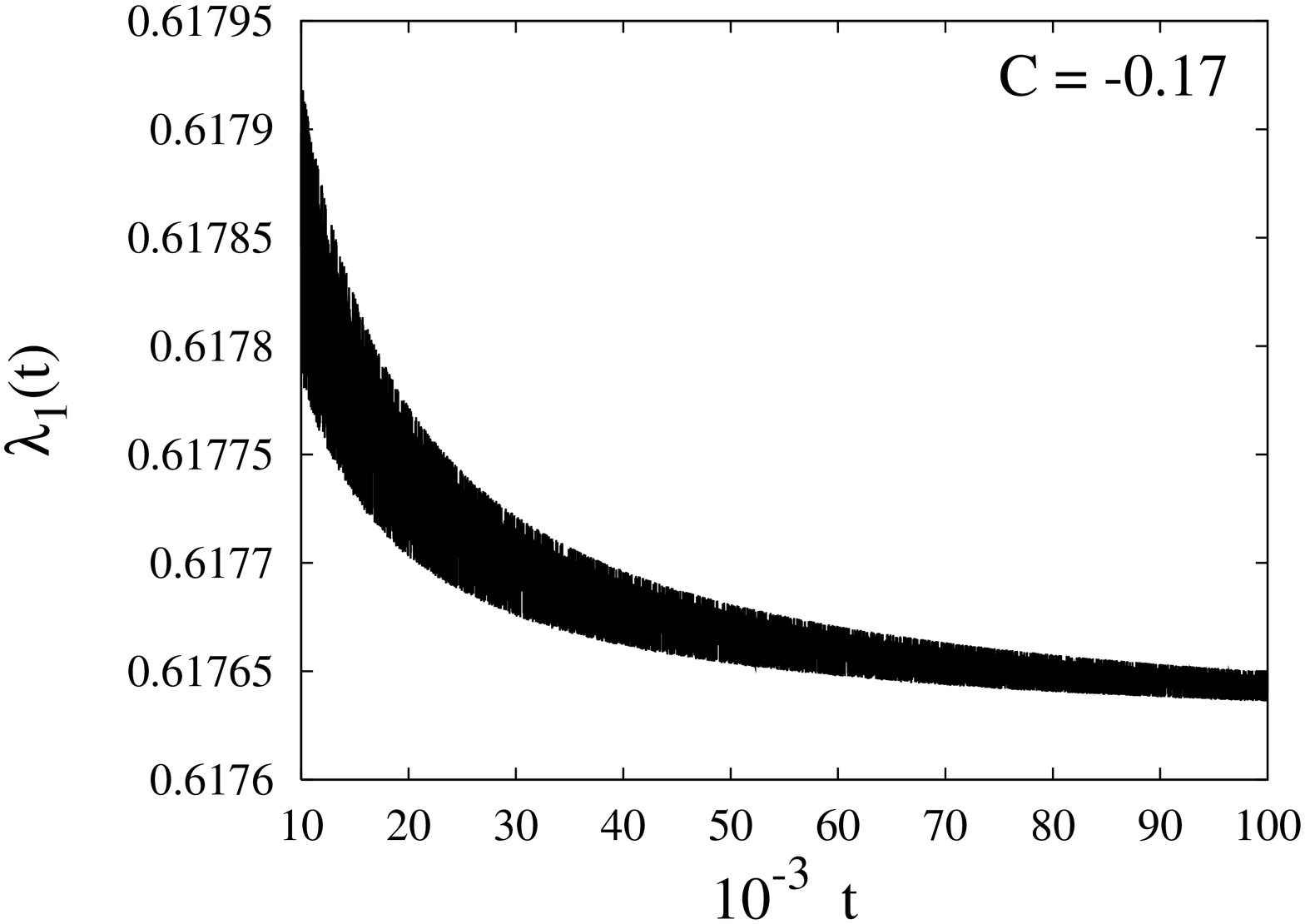}}
\end{minipage}
\begin{minipage}[b]{.48\linewidth}
\centerline{\includegraphics[height=\linewidth,angle=-90]{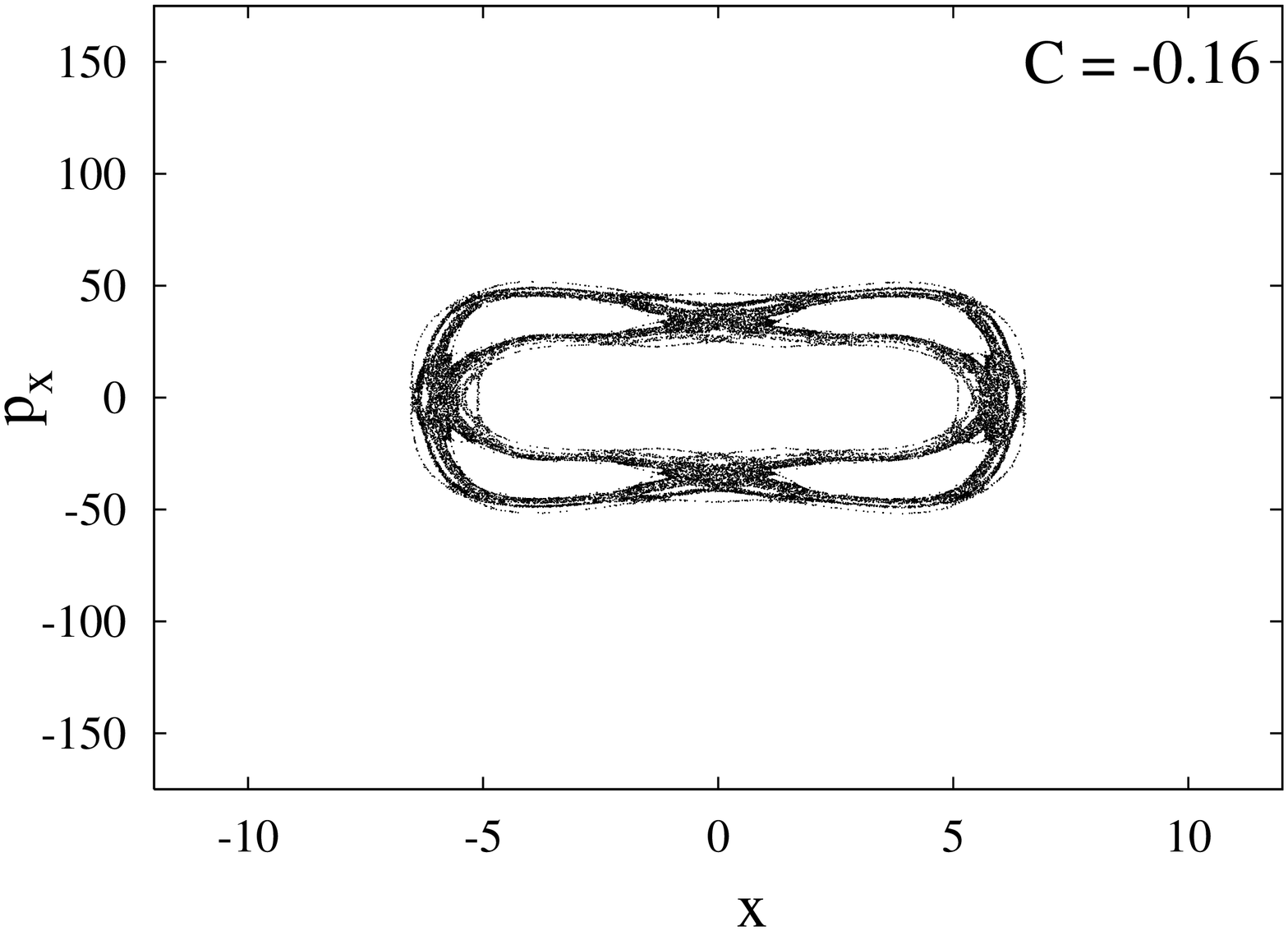}}
\end{minipage}\hfill
\begin{minipage}[b]{.49\linewidth}
\centerline{\includegraphics[height=\linewidth,angle=-90]{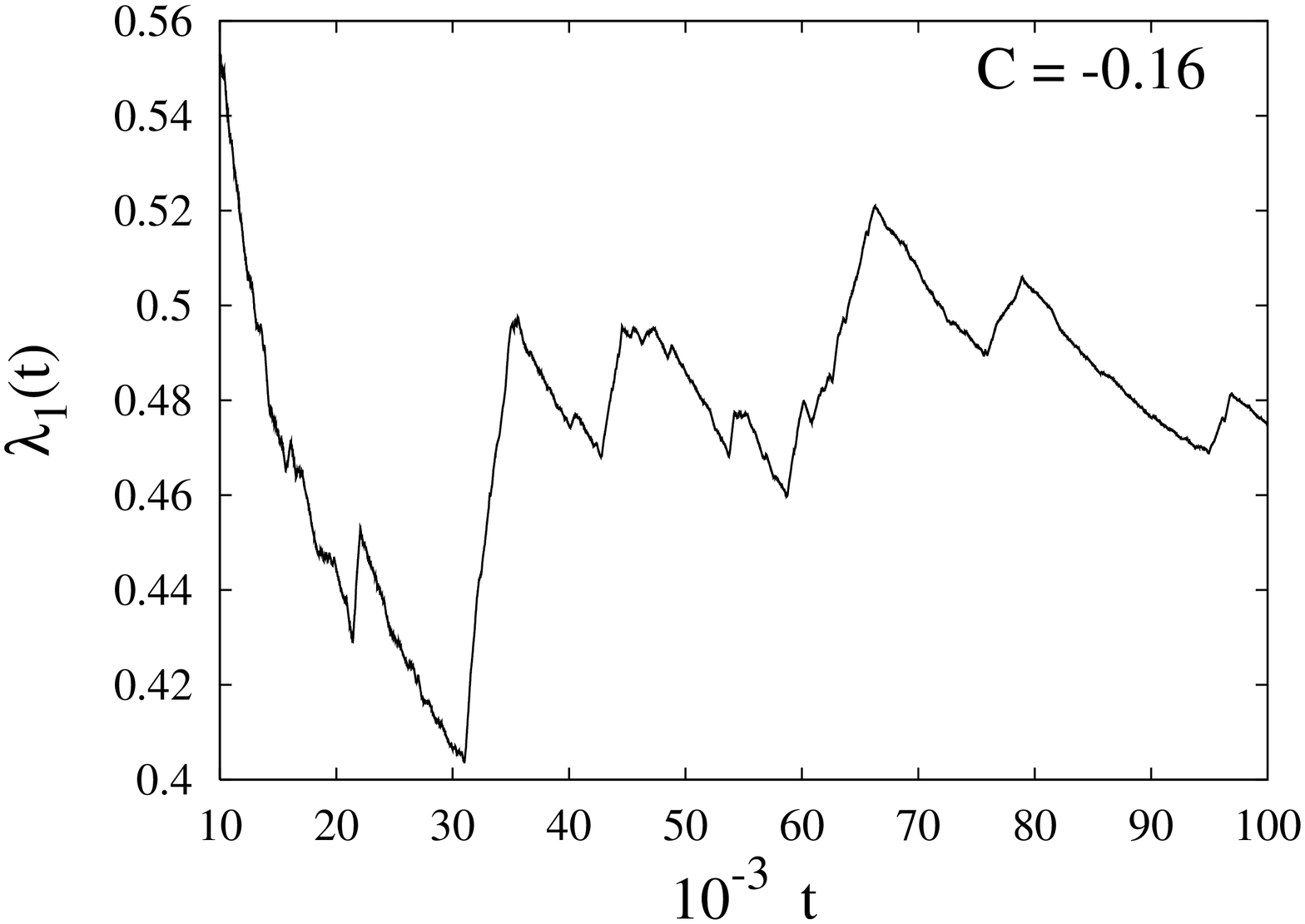}}
\end{minipage}
\begin{minipage}[b]{.49\linewidth}
\centerline{\includegraphics[height=\linewidth,angle=-90]{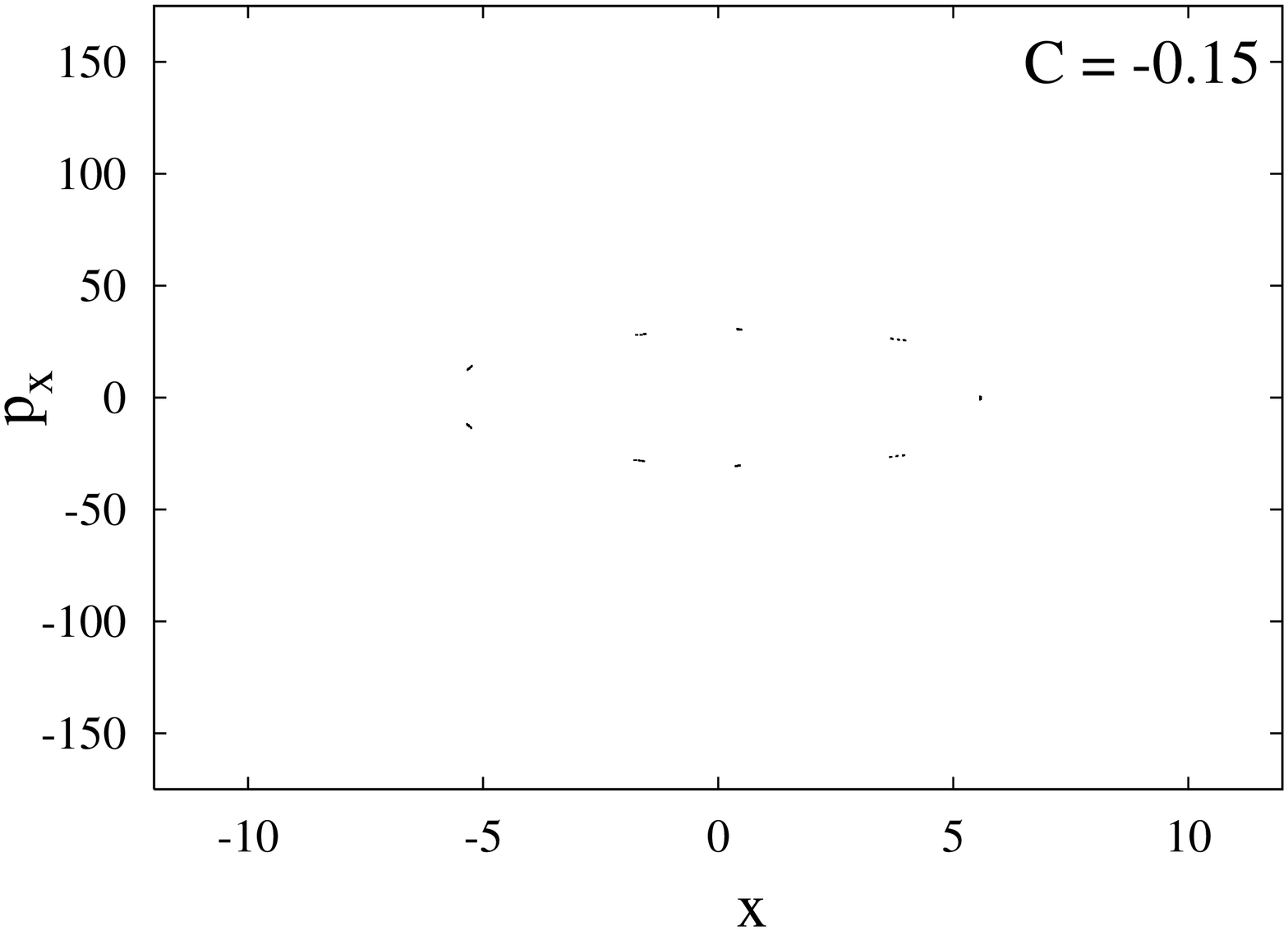}}
\end{minipage}\hfill
\begin{minipage}[b]{.49\linewidth}
\centerline{\includegraphics[height=\linewidth,angle=-90]{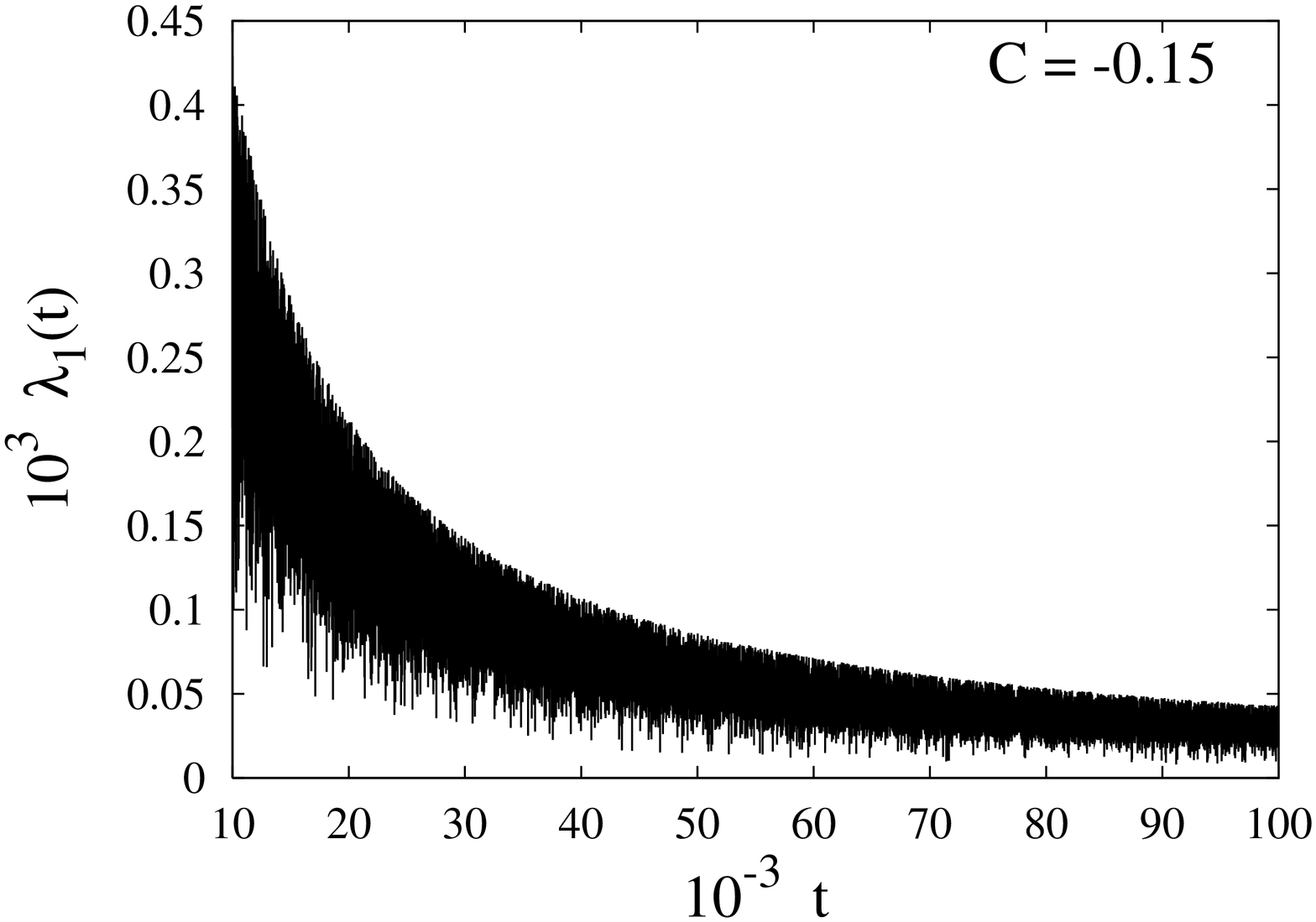}}
\end{minipage}
\caption{\label{f3}(Cont.) Poincar\'e surfaces-of-section (left) and
Lyapunov functions $\lambda_{1}(t)$ from Eq.~(\ref{lyapfunct1_num})
for different values of the coupling constant $C$ of the nonlinear
2D oscillator~(\ref{dodef}).}
\end{figure}

For the coupling constants $C=-0.20$, $C=-0.17$, $C=-0.15$, as
shown in Fig.~\ref{f3}, the Lyapunov functions~(\ref{lyapfunct1_num})
clearly approach limit values in the sense of
Eq.~(\ref{lyaplimit}), which indicates a regular motion each.
The corresponding Poincar\'e surfaces-of-section show that
the respective trajectories cross the plane $y=0$ only in the
neighborhood of previous crossing points in the
$(x,p_{x})$-phase-space plane.
This depicts a quasiperiodic motion, which furnishes---on
the long term---one or more closed curves in the
Poincar\'e map.

In contrast, for coupling constants of $C=-0.21$,
$C=-0.19$, $C=-0.18$, $C=-0.16$, limit values of the respective
Lyapunov functions $\lambda_{1}(t)$ obviously do not exist.
In complete agreement with this, the Poincar\'e surface-of-sections
depicting the crossing points of the plane $y=0$ within
the $(x,p_{x})$-phase-space plane show randomly scattered points.
This reflects the fact that the motions are no longer quasiperiodic,
hence that the respective trajectories are irregular.
\subsection{Circular restricted three-body system}
As a third example, we apply our regularity analysis to
the circular restricted three-body problem\cite{szebe}
of celestial mechanics.
This system describes the motion of a body of negligible mass that
is attracted by two heavy primary bodies.
The primaries are assumed to follow circular orbits around their
common center of mass and, secondly, not to be influenced by the
presence of the third body.
Working with normalized variables in the sideral (fixed) coordinate
system, the equation of motion of the third body contains only one
parameter $\mu$ and is explicitly time dependent\cite{szebe,vela}.
The Hamiltonian of the third body is given by
\begin{equation}\label{ham-crtbp}
H=\onehalf\left(p_{x}^{2}+p_{y}^{2}+p_{z}^{2}\right)+V(x,y,z,t),
\end{equation}
with the time-dependent potential
\begin{eqnarray}\label{pot-crtbp}
&&V(x,y,z,t)=-\frac{\mu}{\rho_{1}}-\frac{1-\mu}{\rho_{2}},\\
\qquad\rho_{1}^{2}&\!=\!\!&{\left[x-(1-\mu)\cos{t}\right]}^{2}+
{\left[y-(1-\mu)\sin{t}\right]}^{2}+z^{2},\nonumber\\
\qquad\rho_{2}^{2}&\!=\!\!&{\left(x+\mu\cos{t}\right)}^{2}+
{\left(y+\mu\sin{t}\right)}^{2}+z^{2}.\nonumber
\end{eqnarray}
The system parameter $\mu$ is determined by the ratio of
the mass of one primary body to the system's total mass.
In the following, we will investigate the orbits of a light third body
(e.g., a comet) in the Sun-Jupiter system with the mass parameter
\begin{displaymath}
\mu=\frac{m_{\text{J}}}{m_{\text{J}}+m_{\text{S}}}\approx 0.0009537,
\end{displaymath}
where $m_{\text{J}}$ and $m_{\text{S}}$ are the masses of
Jupiter and Sun, respectively.
The canonical equations of the third body follow as
\begin{eqnarray}\label{eqm-crtbp}
\dot{x}&=&p_{x},\quad\dot{y}=p_{y},\quad\dot{z}=p_{z},\nonumber\\
\dot{p}_{x}&=&-\mu\frac{x-(1-\mu)\cos{t}}{\rho_{1}^{3}}-
(1-\mu)\frac{x+\mu\cos{t}}{\rho_{2}^{3}},\nonumber\\
\dot{p}_{y}&=&-\mu\frac{y-(1-\mu)\sin{t}}{\rho_{1}^{3}}-
(1-\mu)\frac{y+\mu\sin{t}}{\rho_{2}^{3}},\\
\dot{p}_{z}&=&-\mu\frac{z}{\rho_{1}^{3}}-(1-\mu)\frac{z}{\rho_{2}^{3}}.\nonumber
\end{eqnarray}
Since the Hamiltonian from Eq.~(\ref{ham-crtbp}) is explicitly
time dependent, we must solve the third-order equation~(\ref{adauxeq})
to work out the Lyapunov functions~(\ref{glyapfunct}) in order to
quantify the irregularity of the third body's motion.
For the potential~(\ref{pot-crtbp}), the related
coefficients $g_{1}(t)$ and $g_{2}(t)$ evaluate to
\begin{eqnarray}
g_{1}(t)&=&\frac{4\mu(1-\mu)}{x^{2}+y^{2}+z^{2}}(x\sin{t}-
y\cos{t})\left(\rho_{1}^{-3}-\rho_{2}^{-3}\right),\nonumber\\
g_{2}(t)&=&\frac{-2}{x^{2}+y^{2}+z^{2}}\nonumber\\
&&\quad\times\bigg(\mu\frac{\rho_{1}^{2}-
(1-\mu)(x\cos{t}+y\sin{t}-[1-\mu])}{\rho_{1}^{3}}\nonumber\\
&&\quad\mbox{}+(1-\mu)\frac{\rho_{2}^{2}+
\mu(x\cos{t}+y\sin{t}+\mu)}{\rho_{2}^{3}}\bigg).
\end{eqnarray}
In order to prevent the occurrence of large intermediate
values, the actual numerical calculation of
$\lambda_{1}(t)=-\lambda_{2}(t)-\lambda_{3}(t)$
is performed on the basis of Eqs.~(\ref{adauxeqa}),
(\ref{adauxeqb}) in place of the equivalent Eq.~(\ref{adauxeq}).
\begin{figure}[t]
\begin{minipage}[b]{.49\linewidth}
\centerline{\includegraphics[height=\linewidth,angle=-90]{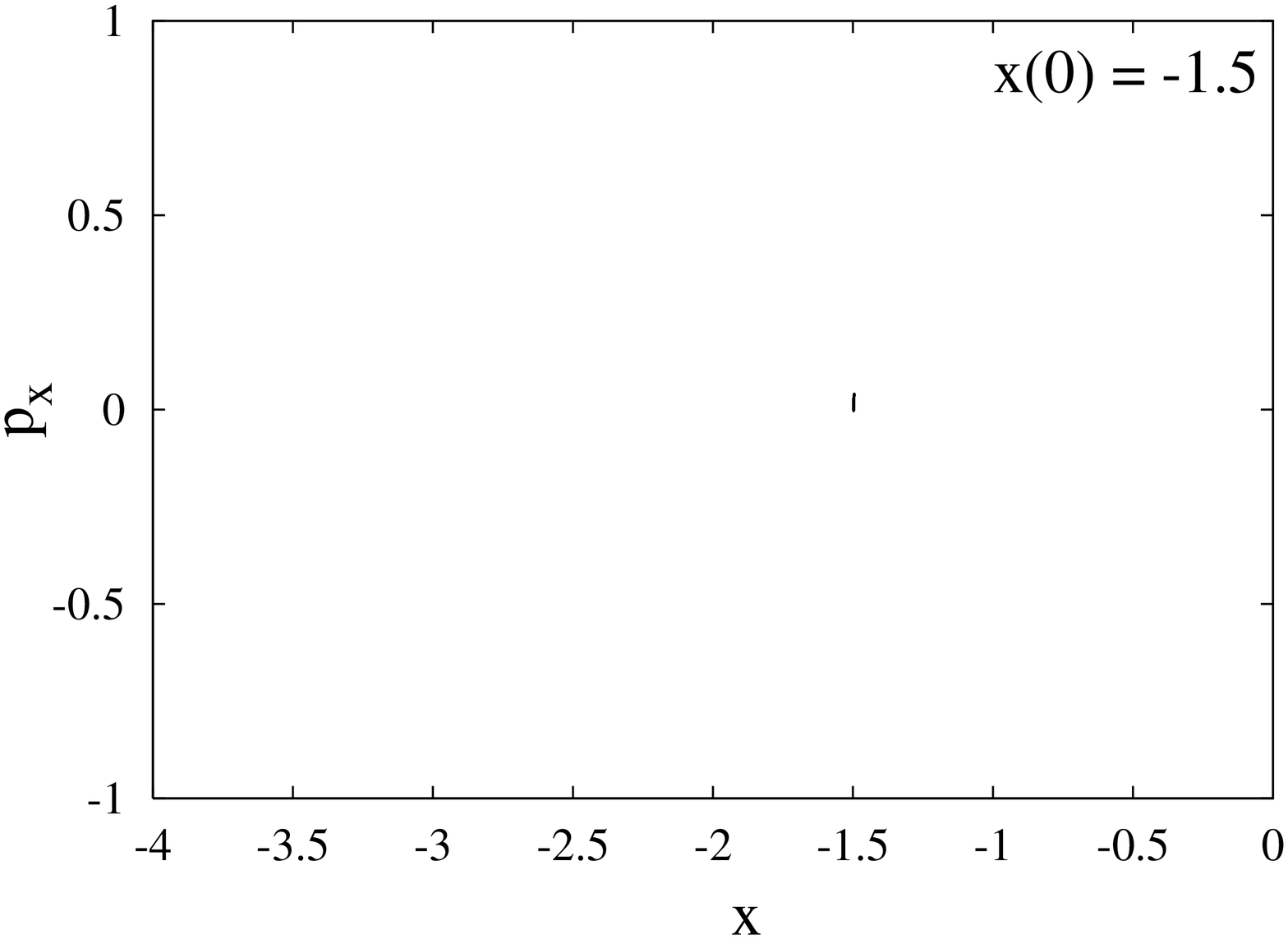}}
\end{minipage}\hfill
\begin{minipage}[b]{.49\linewidth}
\centerline{\includegraphics[height=\linewidth,angle=-90]{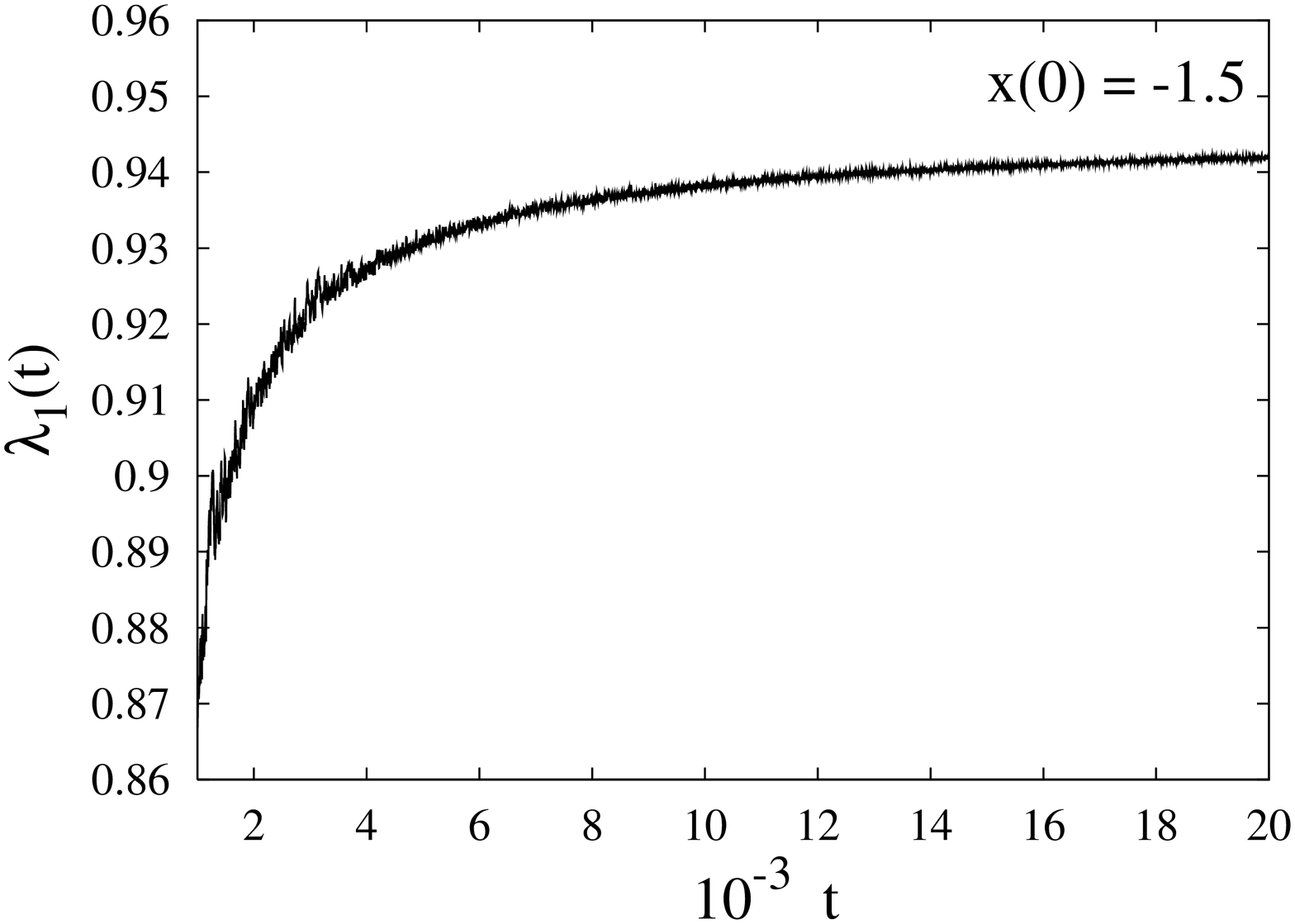}}
\end{minipage}
\begin{minipage}[b]{.49\linewidth}
\centerline{\includegraphics[height=\linewidth,angle=-90]{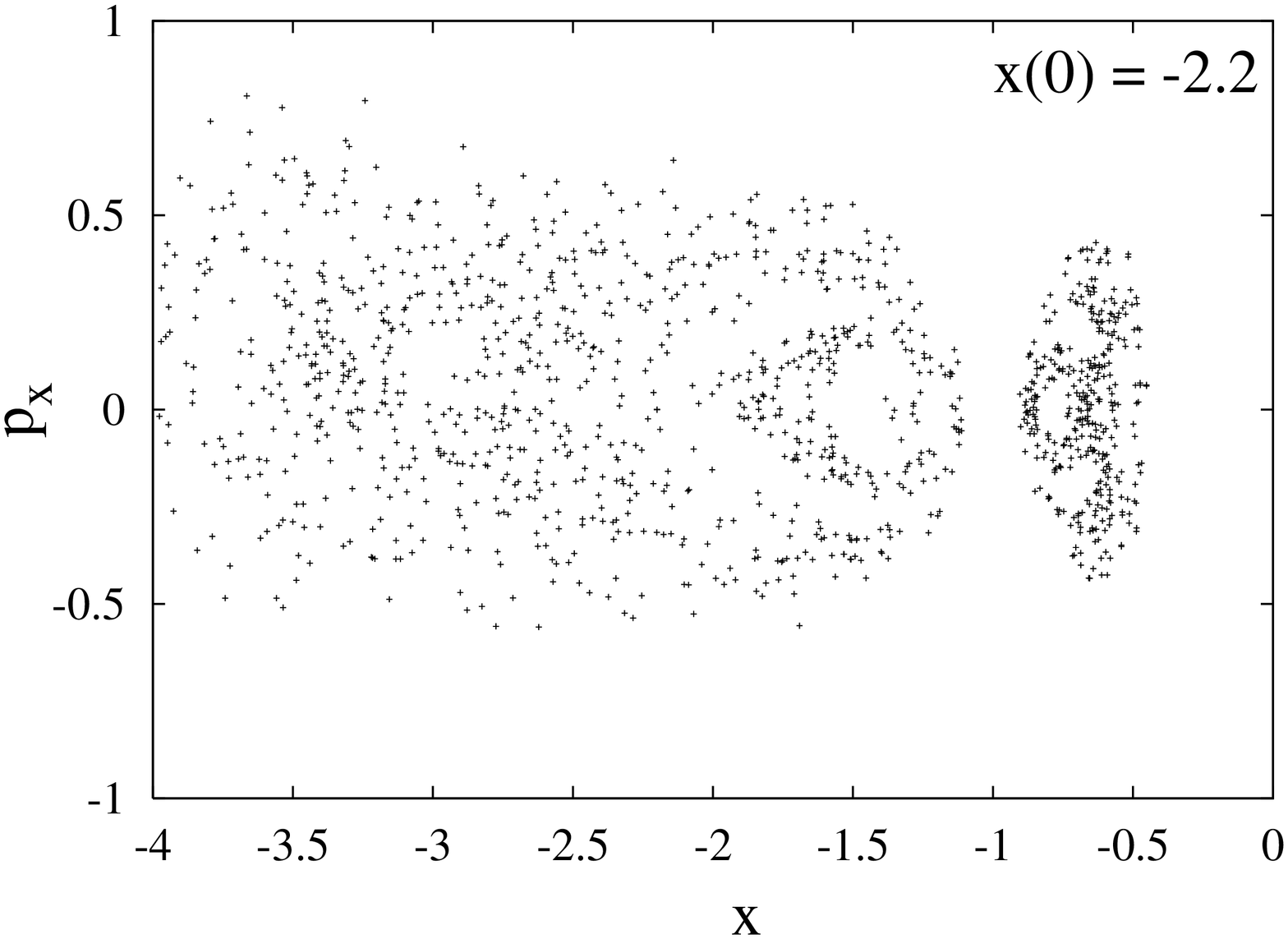}}
\end{minipage}\hfill
\begin{minipage}[b]{.49\linewidth}
\centerline{\includegraphics[height=\linewidth,angle=-90]{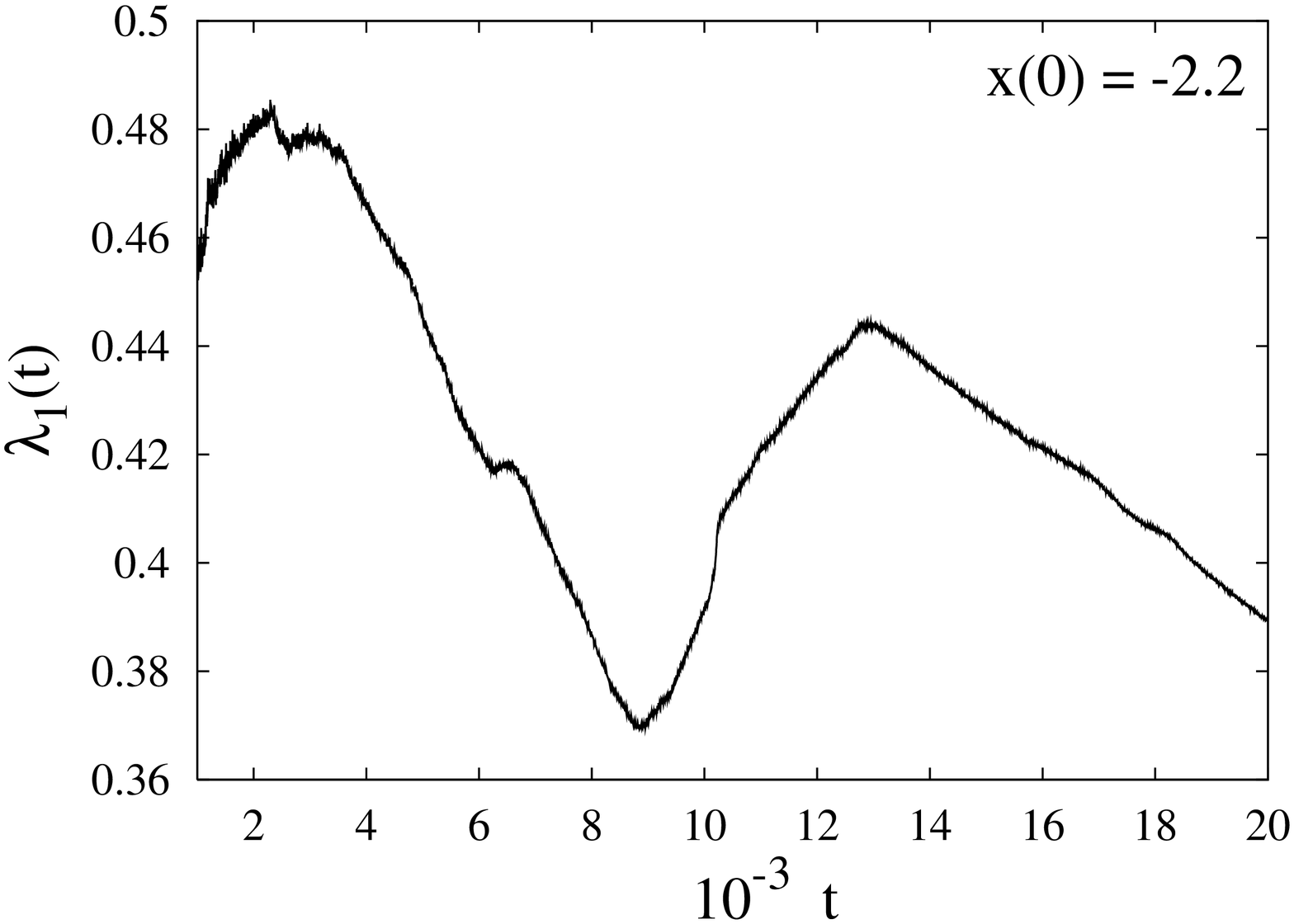}}
\end{minipage}
\caption{\label{f4}Poincar\'e surfaces-of-section (left) and Lyapunov
functions $\lambda_{1}(t)$ from Eq.~(\ref{glyapfunct})
for different initial conditions $x(0)$ of the circular restricted
three-body problem of Eq.~(\ref{eqm-crtbp}) at $E=-1.515$.}
\end{figure}
Figure~\ref{f4} shows the Poincar\'e surface-of-section and the
Lyapunov function $\lambda_{1}(t)$ of Eq.~(\ref{glyapfunct}) for
different initial conditions for $x(0)$.
Note that the corresponding initial conditions are given in the
synodic, i.e., rotating coordinate system by $y(0)=0$,
$p_{x}(0)=v_{x}(0)-y(0)=0$, $p_{y}(0)=v_{y}(0)+x(0)$, where
$v_{y}(0)$ can be obtained from the value of the energy in
the synodic system, i.e., $E=-1.515$ in the example studied
(cf.\ the analysis of Ref.~\cite{vela}).
As has been shown in this paper, the initial value
$x(0)=-1.5$ corresponds to the so called $2:3$ resonance
island, where regular motion of the comet Oterma occurs
with an almost constant frequency.
In the related Poincar\'e surface-of-section this quasicyclic
motion manifests itself as a single spot of finite width.
If the initial value $x(0)$ is chosen outside of this resonance
region, e.g., $x(0)=-2.2$, we observe a nonconverging time
evolution of the Lyapunov function $\lambda_{1}(t)$.
This indicates an irregular motion, which is in accordance
with the corresponding Poincar\'e surface in Fig.~\ref{f4}.
\section{Conclusions}
The method of deducing the degree of irregularity of
Hamiltonian systems from the energy-second-moment map---rather
than from a conventional stability matrix---was
successfully illustrated by means of three examples.
The energy-second-moment map was shown to be represented by
the solution matrix of a linear, homogeneous third-order
equation that is solvable on the top of the complete
set of canonical equations.
In terms of the solutions of this third-order equation,
the Lyapunov functions of the energy-second-moment map
turned out to have simple analytical representations.
The existence of analytical representations of the Lyapunov
functions of the energy-second-moment map simplifies
significantly the irregularity analysis of Hamiltonian systems.
This is the main benefit of our approach and applies to all
systems whose Hamiltonian can be converted into the generic
form $H=\vecp^{2}/2+V(\vecq,t)$.

The conventional stability matrix analysis is based on the
relative motion of infinitesimally separated trajectories.
In contrast, the energy-second-moment-map analysis deals with
the dependence of a {\em single trajectory\/} on its
{\em initial state}, which, in terms of the energy-second moment
vector $\vecs$, is represented by a linear transformation.
From this viewpoint, our approach corresponds to the visual
analysis that is provided by a Poincar\'e surface-of-section.
The latter provides the information on the system's
irregularity through the {\em fractal dimension\/} of the set
of intersections of a {\em single\/} trajectory with the
surface-of-section and likewise does not take into account the
relative motion of neighboring trajectories.
\begin{acknowledgments}
The authors are indebted to C.~Riedel (GSI) for his essential
contributions to this work.
They also acknowledge gratefully the collaboration of S.~Y.~Lee
(University of Indiana) during his sabbatical staying at GSI.
\end{acknowledgments}
\appendix
\section*{Maple worksheet for the symbolic calculation
of Eqs.~(\ref{glyapfunct}) from Eq.~(\ref{wronski})}
\begin{verbatim}
> with(linalg):
> # define matrix Xi(t) [Eq. (9)]
> Xi:=matrix(3,3,[xi1(t),xi2(t),xi3(t),
    diff(xi1(t),t),diff(xi2(t),t),
    diff(xi3(t),t),diff(xi1(t),t$2),
    diff(xi2(t),t$2),diff(xi3(t),t$2)]);
\end{verbatim}
\begin{displaymath}
\Xi:=\left(\begin{array}{ccc}
\xi{1}(t)&\xi{2}(t)&\xi{3}(t)\\
\frac{\d}{\d t}\xi{1}(t)&\frac{\d}{\d t}\xi{2}(t)&
\frac{\d}{\d t}\xi{3}(t)\\
\frac{\d^{2}}{\d t^{2}}\xi{1}(t)&\frac{\d^{2}}{\d t^{2}}\xi{2}(t)&
\frac{\d^{2}}{\d t^{2}}\xi{3}(t)\end{array}\right)
\end{displaymath}
\begin{verbatim}
> # QR decomposition of Xi(t) [Eq. (15)]
> R:=QRdecomp(Xi, Q='q'):
> R:=simplify(R,{det(Xi)=1}):
> # evaluate inverse of R(t)
> Rinv:=inverse(R):
> # evaluate time derivative of R(t)
> Rdot:=matrix(3,3,[diff(R[1,1],t),
      diff(R[1,2],t),diff(R[1,3],t),
      diff(R[2,1],t),diff(R[2,2],t),
      diff(R[2,3],t),diff(R[3,1],t),
      diff(R[3,2],t),diff(R[3,3],t)]):
> # evaluate B(t) [Eq. (16)]
> B:=evalm(Rdot &* Rinv):
> # test trace of B(t)
> TrB:=simplify(B[1,1]+B[2,2]+B[3,3]);
                            TrB := 0
> # evaluate lambda_2(t), lambda_3(t),
> #          lambda_1(t) [Eq. (17)]
> lambda[2]:=int(B[1,1],t)/t;
\end{verbatim}
\begin{displaymath}
\lambda_{2}:=\frac{1}{2t}\ln\left[\xi 1(t)^{2}+
{\left(\frac{\d}{\d t}\xi 1(t)\right)}^{2}+
{\left(\frac{\d^{2}}{\d t^{2}}\xi 1(t)\right)}^{2}\right]
\end{displaymath}
\begin{verbatim}
> lambda[3]:=int(B[3,3],t)/t;
> lambda[1]:=-lambda[2]-lambda[3].
\end{verbatim}

\end{document}